\documentclass[10pt,journal,a4paper]{IEEEtran}
\usepackage{cite}   
\usepackage[caption=false, font=footnotesize]{subfig}
\usepackage[dvips]{graphicx,color}
\usepackage{amsmath,amsfonts,amssymb,mathtools}
\usepackage{enumerate}
\usepackage{times}
\usepackage{tikz}
\usetikzlibrary{arrows,automata}
\usepackage{algpseudocode}
\usepackage{algorithm}

\usepackage{color}

\newtheorem{definition}{Definition}[section]
\newtheorem{lemma}{Lemma}[section]
\newtheorem{theorem}{Theorem}[section]
\newtheorem{corollary}{Corollary}[section]
\newtheorem{remark}{Remark}[section]

\newcommand{\PaperTitle}{Analysis and Optimization of Sparse Random Linear Network Coding for Reliable Multicast Services}

\begin{document}

\title{\PaperTitle}

\author{
	Andrea Tassi, Ioannis Chatzigeorgiou and Daniel E. Lucani
	\thanks{This work is part of the R2D2 project, which is supported by EPSRC under Grant EP/L006251/1, and by the TuneSCode project (No. DFF $1335-00125$) granted by the Danish Council for Independent Research.
	
	A. Tassi was with the School of Computing and Communications, Lancaster University, UK. He is now with the Department of Electrical and Electronic Engineering, University of Bristol, UK (e-mail: a.tassi@bristol.ac.uk).
	
	I. Chatzigeorgiou is with the School of Computing and Communications, Lancaster University, Lancaster, UK (e-mail: i.chatzigeorgiou@lancaster.ac.uk).
	
	D. E. Lucani is with the Department of Electronic Systems, Aalborg University, Aalborg, DK (e-mail: del@es.aau.dk).}
	}

\maketitle

\begin{abstract}
Point-to-multipoint communications are expected to play a pivotal role in next-generation networks. This paper refers to a cellular system transmitting layered multicast services to a multicast group of users. Reliability of communications is ensured via different Random Linear Network Coding (RLNC) techniques. We deal with a fundamental problem: the computational complexity of the RLNC decoder. The higher the number of decoding operations is, the more the user's computational overhead grows and, consequently, the faster the battery of mobile devices drains. By referring to several sparse RLNC techniques, and without any assumption on the implementation of the RLNC decoder in use, we provide an efficient way to characterize the performance of users targeted by ultra-reliable layered multicast services. The proposed modeling allows to efficiently derive the average number of coded packet transmissions needed to recover one or more service layers. We design a convex resource allocation framework that allows to minimize the complexity of the RLNC decoder by jointly optimizing the transmission parameters and the sparsity of the code. The designed optimization framework also ensures service guarantees to predetermined fractions of users. The performance of the proposed optimization framework is then investigated in a LTE-A eMBMS network multicasting H.264/SVC video services.
\end{abstract}


\section{Introduction}\label{sec.Intro}
Among the major novelties likely to be implemented in next-generation networks, there is the possibility of providing services characterized by an availability level of almost $100\%$. In the literature, that emerging kind of services is usually referred to as \emph{ultra-reliable services}~\cite{A}. The ultra-reliable way of conveying services is expected to be greatly useful in a plethora of applications, such as reliable cloud-connectivity, data harvesting from sensors, professional communications~\cite{Ai}. 

Among the possibilities, this paper refers to a system model where a Base Station (BS) transmits, in a multicast fashion, a Point-to-Multipoint (PtM) service to a Multicast Group (MG) of users. In particular, the multicast service is provided in an ultra-reliable way, hence, the service shall be received by predetermined fractions of users, and has to meet target temporal constraints. It is worth noting that the possibility of managing ultra-reliable multicast applications is pivotal, in any Professional Mobile Radio (PMR) standard~\cite{6619579}. Even though classic PMR standards, like Terrestrial Trunked Radio (TETRA) or Association of Public-Safety Communications Officials-Project 25 (APCO P25), refer to ad-hoc communication protocol stacks, the upcoming evolutions of those standards will rely on the 3GPP's Long Term Evolution-Advanced (\mbox{LTE-A}) standard and its extents~\cite{Tetra5G}. As a result, next-generation PMR standards are expected to enable the deployment of PMR systems over pre-existing \mbox{LTE-A} networks.

In this paper, we consider a system model where the base station multicasts a scalable service composed by one \emph{base layer} and multiple \emph{enhancement layers}. The base layer provides a basic reconstruction quality that is gradually improved as one or more enhancement layers are progressively received. Because of the layered nature of the considered multicast service, it is natural to refer to service reliability constraints, which impose that at least a minimum number of users is able to recover predetermined sets of service layers, by a given temporal deadline. The layered service approach has been originally adopted in video communications~\cite{6025326}. However, as discussed in~\cite{A} and~\cite{MetisAL}, the same principle is likely to go beyond the traditional boundaries of multimedia communications and be applied in other fields in order to achieve an \emph{analog-like} service degradation.

Because of the ultra-reliable nature of the considered multicast service, users are required to acknowledge to the base station when they successfully recovered one or more service layers. Even though there exists Automatic Repeat-reQuest (ARQ)~\cite{4441773} and Hybrid ARQ error control protocols~\cite{KiJiKSSc10} suitable for PtM communications, the protocol complexity and the required amount of feedback quickly become intractable as the number of users increases. For these reasons, the reliability of PtM communications is ensured via Application Level-Forward Error Correction (AL-FEC) techniques based on Luby Transform (LT) or low-density parity-check codes. However, as noted in~\cite{6416071}, these kind of codes require large block lengths to operate close to their capacity, and that could potentially be an issue, in the case of multimedia communications. In addition, the most recent evolutions of LT codes~\cite{RaptorQ} usually rely on fixed degree distribution functions and, hence, the code sparsity cannot be optimized on-demand. To this end, in order to mitigate those issues, our system model ensures reliability of multicast communications, via Random Linear Network Coding (RLNC) techniques~\cite{6353397,jsacTassi}.

Given a source message of $k$ source packets to be multicast, the RLNC principle generates and multicasts a stream of coded packets, where each of them is obtained as a linear combination of multiple source packets. A user recovers the source message as soon as it collects a number of linearly independent coded packets that is equal to $k$. RLNC schemes have been used in several wireless settings as a versatile solution for reliable service delivery~\cite{6774596,KODO}. Among the literature contributions, M. Xiao~\textit{et al.}~\cite{RR1} refer to a system model where nodes are connected by a network that can be represented by a Direct Acyclic Graph (DAG); that network consists of one source node and several sinks. In~\cite{RR1}, the RLNC principle takes place at the network layer and allows intermediate nodes to combine several incoming data flows; reliability of coded packet transmissions is ensured via a channel code operating at the physical layer. The size of coded packets and the channel code rate are jointly optimized to minimize the end-to-end delay at the network layer. In addition, multiple resource allocation approaches have been proposed to improve the reliability of layered services via different RLNC implementations~\cite{R1,R2,R3}. In particular,~\cite{R1} considers a multi-hop directed acyclic graph network topology where a scalable service is multicast to multiple receivers. That paper proposes to optimize the communication rate on each link, in order to improve  reliability. Channel erasures are further mitigated via classic FEC techniques. Similarly to~\cite{R1},~\cite{R2} deals with multi-hop network topologies and layered services. However, in that case, reliability of end-to-end communications is improved via a specific implementation of RLNC, which achieves a ladder-shaped global coding matrix. Differently than~\cite{R1} and~\cite{R2},~\cite{R3} applies RLNC to populate a distributed caching system, kept by intermediate network nodes. The communication-ends can take advantage of that while they retrieve the desired scalable service, via a reduced number of Point-to-Point sessions. In contrast to~\cite{RR1,R1,R2,R3}, this paper refers to a typical cellular network topology, where the source node transmits streams of coded packets to a set of users in a multicast fashion. In other words, this paper adopts RLNC to improve reliability over a one-hop broadcast network and not as a way to improve the end-to-end communication throughout across a multi-hop network topology~\cite[Ref.~{[14]}-{[16]}]{jsacTassi} and~\cite[Ref.~{[26]}]{6849990}. 

We observe that the application of RLNC to one-hop broadcast networks has been also discussed in~\cite{RR3} and~\cite{RR2}. In both cases, the broadcasting of a set of source packets is split into multiple stages. During the first stage all the source packets are broadcast by the source node, then, in the following stages, the source node and/or an intermediate relay node broadcast streams of coded packets. Both~\cite{RR3} and~\cite{RR2} focus on different forms of Instantly Decodable Network Codes, which generate coded packets in a deterministic fashion, based on multiple user feedback. As a consequence, we observe that the user uplink traffic can quickly become non-negligible as the number of users increases. Given that we will refer to a system model composed by a source node multicasting services to a multicast group composed by a potentially great number of users, it is not appropriate to refer to the strategies as in~\cite{RR3,RR2}. On the other hand, we will refer to classic decodable RLNC strategies (as in~\cite{6849990}) that are characterized by a significantly smaller user feedback footprint.

Unfortunately, as noted in~\cite{LucaniNetCod,ComplexityGE}, the flip side of the considered RLNC techniques is represented by the complexity of the decoding operations that depends, amongst other code parameters, by the length $k$ of the source message. As noted in~\cite{5061923,5723187}, the decoding complexity problem can be partially mitigated by the systematic implementation of RLNC (SRLNC). However, in case of poor propagation conditions, the performance of SRLNC coincides with that of RLNC~\cite{ICicc2015}. Obviously, the more the decoding complexity grows, the more the processing footprint increases and, hence, the battery of mobile devices discharges. For these reasons, this paper addresses the following fundamental question: \emph{Is there a way to minimize the RLNC decoding complexity of ultra-reliable layered multicast communications without altering the decoder currently onboard mobile devices?}

We will answer the aforementioned research question by referring to multiple sparse RLNC techniques. As will be clear in the following section, let us intuitively define the \emph{sparsity} of the code as the number of source packets that on average are involved in the generation process of each coded packet~\cite{LucaniNetCod}. To the best of our knowledge, the general expression of the decoding complexity as a function of the source message length and the sparsity is unknown. However, the decoding complexity decreases as the source message gets shorter~\cite{ComplexityGE} and/or the sparsity increases~\cite{LucaniNetCod}. Intuitively, as the sparsity increases, the information content of each coded packet decreases. Hence, the average number of coded packets needed to recover a source message increases as the sparsity grows~\cite{LucaniNetCod}. That leads us to further refine our research question as follows: \emph{Are there any optimized sparse RLNC strategies ensuring: (i) a reduced decoding complexity, and (ii) a recovery of the source message with an average number of coded packet transmissions, which is close or equal to that provided by non-sparse RLNC techniques?}

The first contribution of the paper is that of providing an efficient performance modeling of sparse non-systematic and systematic RLNC techniques via a \emph{unified} theoretical framework. In particular, in Section~\ref{sec.SM}, we characterize the user performance in terms of the average number of coded packet transmissions needed to recover a given service layer. It is well known in the literature that an exact expression for the aforementioned performance index is unknown~\cite{5963470,5978939,RSA:RSA1,Cooper2000}. That is caused by the lack of an analytical formulation of the probability of generating a full-rank sparse random matrix over a finite field~\cite{Cooper2000}. In order to mitigate the aforementioned issue, X. Li~\textit{et al.}~\cite{5963470,5978939}, proposed a pioneering approach for upper-bounding and lower-bounding the probability of generating at random a sparse non-singular random matrix, based on the zero pattern of the random matrix. Unfortunately, the validity of the resulting bounds has been proven only for large finite fields. Apart from that, those bounds cannot be efficiently incorporated into an optimization model meant to be solved on-demand, before starting the transmission of a service. In fact, the bound expressions involve nested sums where each term is a product of several binomial coefficients, which could not be practically derivable, in the case of large source message lengths (Section~\ref{sec.SM}). Furthermore, it is also not straightforward to formally prove the convexity of the bounds as in~\cite{5963470,5978939}, because their definitions involve several non-differentiable points.

For these reasons, we rely on the results presented in~\cite{RSA:RSA1} and extended in~\cite{Cooper2000}. However, in~\cite{RSA:RSA1,Cooper2000}, authors only provide a lower-bound of the probability that a sparse $(t+1) \times k$ matrix is full-rank, given that the first $t$ rows are linearly independent, for $0 \leq t \leq (k-1)$. It is worth mentioning, that the aforementioned result was provided without referring to any communication system or coding strategy. By building upon that result, we provide an upper-bound for the average number of coded packet transmissions needed to recover a service layer, via an Absorbing Markov Chain (AMC) with reduced complexity. In particular, Section~\ref{subsec.AMC} will show how our performance modeling does not involve any explicit matrix inversion, which is a common and computationally costly step in AMC-based analysis. As will be clear in the following sections, that desirable feature is achieved because of: (i) the nature of the aforementioned probability lower-bound and, (ii) the way we defined the states of the proposed AMC model.

The second contribution of the paper is made in Section~\ref{sec.Opt}, where we answer to our research question by building upon an efficient user performance characterization and proposing a resource allocation framework for ultra-reliable layered multicast services. The proposed framework aims to maximize the code sparsity associated to each service layer, and hence, the overall decoding complexity is minimized. The optimization goal is fulfilled by a joint optimization of both the code sparsity and the  Modulation and Coding Schemes (MCSs) used for multicasting each service layer. In addition, given the layered nature of the transmitted services, the optimization constraints ensure that the desired number of service layers are recovered by predetermined fractions of users, with an average number of coded packet transmissions that is smaller than or equal to a target value. We prove that the proposed resource allocation framework is convex and can be easily solved. Finally, we remark that the proposed resource allocation framework applies for several sparse RLNC techniques, in a complete RLNC decoder-agnostic fashion. 

Even though our analysis deals with a generic cellular system model, Section~\ref{sec.AR} inspects the effectiveness of the proposed optimized sparse RLNC techniques by referring to a \mbox{LTE-A} communication network. We chose that particular communication standard for two main reasons: (i) LTE-A is likely to play a leading role in the early-stage deployment of next-generation networks~\cite{ericsson}, and (ii) LTE-A provides the support to handle PtM communications at the radio access and core network level, by means of the evolved Multimedia Broadcast Multicast Service (eMBMS) framework~\cite{sesia2011lte}. In the proposed performance investigation, we refer to a MG targeted by non-real time multimedia multicast services compressed according to the widely used H.264 video encoding standard. In particular, we referred to the scalable extension of H.264, called Scalable Video Coding (H.264/SVC)~\cite{h264}. In line with the considered system model, an H.264/SVC video stream consists of several layers such that the enhancement layers improve the reconstruction quality provided by the base video layer. Finally, Section~\ref{sec.cl} summarizes the main findings of the paper.

\section{System Model and Performance Characterisation}\label{sec.SM}
We consider a one-hop broadcast communication system, which is composed by one source node and a MG of $U$ users (hereafter called \emph{multicast users}). In order to improve the reliability of PtM communications, the source transmits data streams encoded according to the RLNC principle. As a consequence, the source node transmits streams of network-coded packets (henceforth referred to as \emph{coded packets}) to the MG. For the sake of generality, we assume that the transmission of a PtM communication occurs over a set of orthogonal broadcast erasure subchannels. Each subchannel consists of basic resource allocation units called \emph{resource blocks}.

As mentioned in the previous section, our main goal is to design a general optimized service-provisioning paradigm for ultra-reliable multicast services, with a reduced decoding computational complexity. The following section will also clarify that the proposed theoretical modeling (Section~\ref{subsec.AMC}) and the resource allocation procedure (Section~\ref{sec.Opt}) are easily applicable to any cellular system capable of multicasting multiple data streams at the same time. However, in order to effectively map user Quality of Service (QoS) constraints onto typical system performance metrics (e.g., delay, packet error rate, etc.), we will refer to an OFDM-based multicarrier communication system.

In the considered physical layer, the downlink phase is organized in radio frames. Resource blocks forming each subchannel are transmitted in one or more radio frames. Each frame can be modeled as a frequency $\times$ time structure where the frequency and time domains are discretized into OFDM subcarriers and OFDM symbols, respectively. Each resource block occupies a fixed time interval ($\Hat{\tau}_{\mathrm{RB}}$) and frequency band, i.e., each resource block spans a fixed number of OFDM symbols and OFDM subcarriers. Since multicast users may experience heterogeneous propagation conditions, user diversity is exploited by assuming that the subcarriers used in a resource block are selected at random among all the available ones~\cite{6148193}. We also assume that users are static or characterized by low mobility, hence, the user channel conditions are considered constant within a resource block.

Each coded packet is always mapped onto one resource block and transmitted by means of a specific MCS that is identified by an index, which can take $M$ possible values. We denote by $p_{u}(m)$ the Packet Error Rate (PER) experienced by a multicast user $u$, and by $r(m)$ the number of information bits carried by one resource block, when the MCS with index $m$ is in use. Let us consider two MCSs with indexes $a$ and $b$, where $a < b$. In our system model we assume that the MCS with index $a$ is characterized by a smaller modulation order and/or a lower channel code rate than $b$. For the same user propagation conditions, we have $p_{u}(a) \leq p_{u}(b)$ and $r(a) < r(b)$. We also refer to a system where all the resource blocks belonging to the same subchannel shall adopt the same MCS. Coded packets associated with a PtM data service are transmitted via one or more broadcast erasure subchannels.

The source node transmits to the MG a \emph{layered scalable} service consisting of one basic layer and $L-1$ enhancement layers. Each layer is characterized by different priority levels. The basic layer (also referred to as ``layer $1$'') owns the highest priority, which decreases in the case of the enhancement layers (layers $2, \ldots, L$). In particular, layer $L$ is characterized by the lowest priority. Because of that, it is natural to define the level of QoS achieved by a multicast user as the number of consecutive message layers, starting from the base layer, that can be recovered. Hence, a user shall achieve the QoS level $\ell$, if \emph{all} the layers $1, \ldots, \ell$ are successfully recovered. For instance, if a user successfully recovers message layers $\{1, 2, \ldots, \ell, \ell + 2, \ell + 3, \ldots, L\}$ then layers $2$ to $\ell$ improve the information provided by layer $1$. In that case, the QoS level achieved is equal to $\ell$, and layers $\ell + 2, \ldots, L$ do not provide any QoS improvements, as layer $\ell + 1$ has not been received.

The considered multi-layer principle has been originally designed for video compression standards. In the case of H.264/SVC~\cite{h264}, it is possible to achieve different kinds of video scalability~\cite{6025326}. With the spatial scalability, the video frame resolution is gradually increased by each layer with the purpose to fit screens with different capabilities. In that case, the content provided by layer $1$ allows a user, for instance, to  recover a $352\times288$ px video stream. By following the same train of toughs, the spatial resolution can be boosted to $720\times480$ px and $1920\times1080$ px, by means of layers 1 and 2, and layers 1 to 3, respectively. It is worth mentioning that our analysis is generic enough to be applied to any layered scalable service that follows the previously mentioned hierarchical structure. It is beyond the scope of the paper to provide analytical and optimization frameworks dealing with the compression strategy used to generate a scalable service. For these reasons, the proposed analysis has been made independent of the way service layers are generated and the nature of the adopted service scalability.

As suggested in~\cite{jsacTassi,6849990}, we model the transmitted service as a stream of information messages of the same size. The scalable nature of the service is reflected on each message. In particular, each message consists of $L$ layers, where layer $\ell$ is a sequence of $b_\ell$ bits. We remark that coded packets associated with different message layers are transmitted by different subchannels. Therefore, the total number of occupied subchannels \mbox{is $L$}. In the rest of the paper, we will provide an analytical framework suitable for optimizing the transmission of each message and, hence, of the whole layered service.

\begin{figure}[tb]
\centering
\includegraphics[width=0.7\columnwidth]{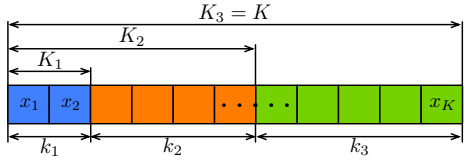}
\caption{Layered source message, in the case of $L = 3$.}
\label{fig.msg}
\end{figure}

Each layered message $\mathbf{x} = \{x_1, \ldots, x_K\}$ consists of $K$ source packets, as shown in Fig.~\ref{fig.msg} for a $3$-layer message. In particular, layer $\ell$ of $\mathbf{x}$ is defined by a fixed number $k_{\ell}$ of source packets, implying that $K = \sum_{\ell = 1}^L k_\ell$. If the MCS adopted by the subchannel delivering coded packets of service layer $\ell$ is $m_\ell$, the number of bits carried by each resource block will be equal to $r(m_\ell)$. Hence, we define $k_{\ell} = \left\lceil b_\ell/r(m_\ell) \right\rceil$. Without loss of generality we assume that the first source packets of $\mathbf{x}$ belong to the base layer ($\ell = 1$), and are progressively followed by packets defining the enhancement layers ($\ell = 2, \ldots, L$).

In the remaining part of the paper, we will characterize the performance of different network coding strategies. It will also become clear how the selection of MCS scheme and sparsity associated with each message layer can be jointly optimized.

\subsection{Random Linear Network Coding Background}\label{subsec.NCbac}
Let $K_\ell = \sum_{t=1}^{\ell}k_t$ be the number of source packets forming the first $\ell$ layers of a source message. In the classic implementation of RLNC, the source node linearly combines source packets $\{x_i\}_{i = K_{\ell-1}+1}^{K_\ell}$ forming message layer $\ell$, in order to generate a stream $\{y_j\}_{j = 1}^{n_\ell}$ of $n_\ell$ coded packets, where $y_j = \sum_{i = K_{\ell-1}+1}^{K_{\ell}}c_{j,i}\cdot x_i$. Each coding coefficient $c_{j,i}$ is uniformly selected at random over a finite field $\mathrm{GF}(q)$ of size $q$. The coding coefficients associated with $y_j$ define the \emph{coding vector} $\mathbf{c}_j = (c_{j,K_{\ell-1}+1}, \ldots, c_{j,K_{\ell}})$. Since each coding coefficient is obtained by the same Pseudo-Random Number Generator (PRNG), modern NC implementations are keen on representing $\mathbf{c}_j$ by the PRNG seed used to compute the first coding vector component $c_{j,K_{\ell-1}+1}$. The seed is transmitted along with the correspondent coded packet. Since each user is equipped by the same PRNG, it can incrementally recompute all the coding vector components, starting from the first one~\cite{6353397,6849990}. The RLNC encoding process is then repeated for each message layer $\ell = 1, \ldots, L$. A multicast user can recover the source message layer $\ell$, if it successfully receives $k_\ell$ linearly independent coded packets associated with that message layer.

Unlike classic RLNC, a coded packet stream obtained by SRLNC associated with layer $\ell$ generates $k_\ell$ \emph{systematic packets} and one or more coded packets. The systematic packets are identical to the source packets $\{x_i\}_{i = K_{\ell-1}+1}^{K_\ell}$, while the coded packets are obtained as in the classic RLNC case. For the sake of the analysis, we define the coding vector associated with systematic packet $i$ as a vector where: (i) the $i$-th component is equal to $1$, and (ii) all the remaining components are equal to $0$. For clarity, we will refer to a coding vector related to a systematic packet as \emph{degenerate} coding vector in the rest of the paper. In our system model, we assume that users acknowledge to the source node, over a fully reliable channel, the successful recovery of a layer. Furthermore, the source node transmits a message layer until a predetermined fraction of multicast users has recovered it. Obviously, as will become clear in Section~\ref{sec.Opt}, the transmission of each layer shall meet a temporal constraint. 

The sparse versions of both the classic (S-RLNC) and systematic implementation of RLNC (S-SRLNC) are obtained as follows. Each component $c_{j,i}$ of a non-degenerate coding vector associated with source message layer $\ell$ is independently and identically distributed as follows~\cite{5978939}:
\begin{equation}\label{eq.pl}
  \mathrm{Pr}\left(c_{j,i} = v\right) = \left\{ 
  \begin{array}{l l}
  	p_\ell & \quad \text{if $v = 0$}\\
    \displaystyle\frac{1-p_\ell}{q-1} & \quad \text{if $v \in \textrm{GF}(q) \setminus \{0\} $}\\
  \end{array} \right.
\end{equation}
where $p_\ell$, for $0 < p_\ell < 1$, is the probability of having \mbox{$c_{j,i} = 0$}. The event $c_{j,i} \neq 0$ occurs with probability $1-p_\ell$. We remark that the average number of source packets involved in the generation of a non-degenerate coded packet, i.e., the \emph{sparsity} of the code, can be controlled by tuning the value of $p_\ell$, for any $\ell = 1, \ldots, L$.

Since coding vectors are generated at random, there is the possibility of generating coding vectors where each coding coefficient is equal to $0$. From a system implementation perspective, all-zero coded packets should be discarded and not transmitted. On the other hand, in the literature dealing with the performance characterization of RLNC, it is common to include the transmission of all-zero coded packets~\cite{4459059,6991518}. In that way, the performance modeling is tractable  and keeps a higher degree of generality. The same principle is adopted in this and the following sections. However, Section~\ref{subsec.ass} will show how the proposed analytical modeling can be applied to a practical communication system where all-zero coded packets are not transmitted.

In order to establish a link between the coding schemes presented in~\cite{jsacTassi} and those discussed in this paper, the following sections will deal with the Non-Overlapping Window (NOW-RLNC) and the systematic NOW-RLNC strategies. We observe that the exact performance model of the Expanding Window RLNC (EW-RLNC) strategy is unknown, even for the non-sparse case. In fact,~\cite{jsacTassi} proposes an upper-bound to the probability of recovering a source message, when the EW-RLNC is used. Since the reasoning behind that bound relies on a well-known result of classic non-sparse RLNC~\cite{5634159}, its extension to the sparse case is not trivial. For these reasons, the sparse implementation of EW-RLNC is still an open research issue.

\subsection{Markovian Modelling for Delay Performance}\label{subsec.AMC}
In this paper, user performance will be expressed in terms of the average number of coded packet transmissions after which a user $u$ achieves a predetermined QoS level. For this reason, in the remainder of the section, we focus on user $u$ and model the recovery of message layer $\ell$ as a Markovian process. In particular, the user decoding process is modeled via an AMC.

Let $\mathbf{C}_u$ be a matrix associated with the user $u$ consisting of $k_\ell$ columns and variable number of rows. As user $u$ successfully receives a coded packet associated with layer $\ell$, the corresponding coding vector is extracted and added, as a new row, into matrix $\mathbf{C}_u$. Assume $u$ already received $n_\ell \geq k_\ell$ coded packets, i.e., $\mathbf{C}_u$ is a $n_\ell \times k_\ell$ matrix. User $u$ recovers layer $\ell$ when the rank of $\mathbf{C}_u$, denoted by $\mathrm{rank}(\mathbf{C}_u)$, is equal to $k_\ell$ or equivalently when the defect of the matrix, defined as $\mathrm{def}(\mathbf{C}_u) = k_\ell - \mathrm{rank}(\mathbf{C}_u)$, is zero. For these reasons, we define a state of the user AMC as follows.
\begin{definition}\label{d.state}
The AMC associated with user $u$ and message layer $\ell$ is in state $s^{(u,\ell)}_{i}$, if $\mathrm{def}(\mathbf{C}_u) = i$, for $i = 0, \ldots, k_\ell$.
\end{definition}

At first, when user $u$ has not received any coded packet or coded packets associated with zero-coding vectors, the defect of $\mathbf{C}_u$ is $k_\ell$, and hence, the AMC is in state $s^{(u,\ell)}_{k_\ell}$. The defect progressively decreases, i.e., the index of the AMC state decreases, as new linearly independent coded packets are received. As a consequence, in the case of layer $\ell$, we have that the AMC consists of $k_\ell+1$ states. Furthermore, in order to define the probability transition matrix of the user AMC, we summarize here the proof of the following lemma, presented in~\cite[Theorem 6.3]{RSA:RSA1}. 
\begin{lemma}[{\hspace*{-1.3mm}\cite[Theorem~6.3]{RSA:RSA1}}]
Assume that matrix $\mathbf{C}_u$ consists of $(t+1) \times k_\ell$ elements, for $0 < t \leq (k_\ell-1)$, and assume that $t$ out of $t+1$ rows are linearly independent. The probability $\mathrm{P}_{\ell,t}$ that matrix $\mathbf{C}_u$ is not full-rank admits the following upper-bound:
\begin{equation}
\mathrm{P}_{\ell,t} \leq \left[\max\left(p_\ell, \frac{1-p_\ell}{q-1}\right)\right]^{k_\ell - t}.\label{eq.LB}
\end{equation}\label{l.singular.b}
\end{lemma}
\begin{IEEEproof}
Without loss of generality, assume that the first $t$ rows of $\mathbf{C}_u$, denoted by $\mathbf{C}_{u,1}, \ldots, \mathbf{C}_{u,t}$, are linearly independent. By resorting to basic row-wise operations, it is possible to transform $\mathbf{C}_u$ such that the first $t$ rows and columns of $\mathbf{C}_u$ define the $t \times t$ identity matrix. Consequently, the first $t$ rows of the transformed $\mathbf{C}_u$ generate the same vector space defined by $\mathbf{C}_{u,1}, \ldots, \mathbf{C}_{u,t}$. The probability that $\mathbf{C}_u$ is not full-rank entirely depends on the last $k_\ell - t$ components of the last row $\mathbf{C}_{u,t+1}$ of $\mathbf{C}_u$. Hence, the probability that $\mathbf{C}_{u,t+1}$ does not belong to the vector space defined by $\mathbf{C}_{u,1}, \ldots, \mathbf{C}_{u,t}$ is at least $1-\max\left(p_\ell, \frac{1-p_\ell}{q-1}\right)^{k_\ell - t}$. That completes the proof.
\end{IEEEproof}

Because of~\eqref{eq.pl}, the exact QoS characterization is a challenging task~\cite{5978939}. In particular, to the best of our knowledge, the exact expression of $\mathrm{P}_{\ell,t}$ is not known. In the rest of the paper, owing to the lack of the exact expression of $\mathrm{P}_{\ell,t}$, we use~\eqref{eq.LB} to approximate $\mathrm{P}_{\ell,t}$, that is
\begin{equation}
\mathrm{P}_{\ell,t} \cong \left[\max\left(p_\ell, \frac{1-p_\ell}{q-1}\right)\right]^{k_\ell - t}.\label{eq.LBapp}
\end{equation}
The following remark is immediate from~\eqref{eq.LB} and~\eqref{eq.LBapp}.

\begin{remark}\label{rem.1}
If $p_\ell=q^{-1}$, each non-degenerate coding vector is equiprobable, for a given value of $k_\ell$. Hence, a coding vector belongs to the vector space generated by $t$ linearly independent coding vectors with probability $\mathrm{P}_{\ell,t} = q^t/q^k_\ell$. 
This result has been discussed in the literature~\cite{5634159} but is clearly not applicable to the sparse case, in contrast to \eqref{eq.LBapp}. It is worth mentioning that the considered approximation~\eqref{eq.LBapp} collapses to the exact expression of $\mathrm{P}_{\ell,t}$ and, hence, the relation $\mathrm{P}_{\ell,t} = \left[\max\left(p_\ell, (1-p_\ell)/(q-1)\right)\right]^{k_\ell - t} = q^t/q^{k_\ell}$ holds, for $p_\ell=q^{-1}$.
\end{remark}

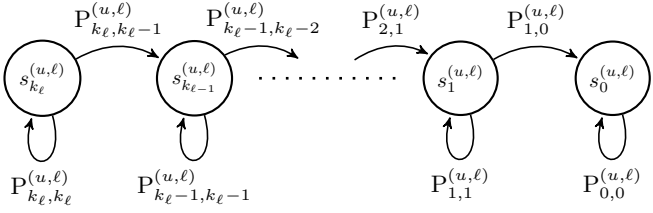
\begin{figure}[tb]
	\begin{center}
\begin{tikzpicture}[->,>=stealth',shorten >=1pt,auto,node distance=4.5cm,semithick]
\tikzstyle{every state}=[fill=white,draw=black,thick,text=black,scale=0.8]
\node[state]         (A) at (-6.5,0) {$s_{k_\ell}^{(u,\ell)}$};
\node[fill=white,draw=black,thick,text=black,scale=0.78,circle]         (B) at (-4.5,0) {$s_{k_{\ell-1}}^{(u,\ell)}$};
\node[fill=white,draw=white,thick,text=black,scale=0.8]         (Bf) at (-3,0.2) {};
\node[fill=white,draw=white,thick,text=black,scale=0.8]         (Cf) at (-2.5,0.2) {};
\draw [-, loosely dotted, line width=1pt] (-3.65,0) -- (-1.75,0);
\node[state]         (C) at (-1,0) {$s_{1}^{(u,\ell)}$};
\node[state]         (D) at (1,0) {$s_{0}^{(u,\ell)}$};

\path (A) edge  [loop below] node {\small{$\mathrm{P}_{k_\ell,k_\ell}^{(u,\ell)}$}} (A);
\path (A) edge  [bend left=30] node[above] {\small{$\mathrm{P}_{k_\ell,k_\ell-1}^{(u,\ell)}$}} (B);

\path (B) edge  [loop below] node {\small{$\mathrm{P}_{k_\ell-1,k_\ell-1}^{(u,\ell)}$}} (B);
\path (B) edge  [bend left=30] node[above] {\small{$\mathrm{P}_{k_\ell-1,k_\ell-2}^{(u,\ell)}$}} (Bf);

\path (Cf) edge  [bend left=30] node[above] {\small{$\mathrm{P}_{2,1}^{(u,\ell)}$}} (C);

\path (C) edge  [loop below] node {\small{$\mathrm{P}_{1,1}^{(u,\ell)}$}} (C);
\path (C) edge  [bend left=30] node[above] {\small{$\mathrm{P}_{1,0}^{(u,\ell)}$}} (D);

\path (D) edge  [loop below] node {\small{$\mathrm{P}_{0,0}^{(u,\ell)}$}} (D);

\end{tikzpicture}
	\end{center}
\vspace{-3mm}
\caption{State transition diagram for the AMC associated with user $u$ and message layer $\ell$.}
	\label{fig:MarkovChain}
\end{figure}

From~\eqref{eq.LBapp}, the transition probability matrix describing the AMC associated with user $u$ and message layer $\ell$ can be derived by the following lemma.
\begin{lemma}\label{l.transition}
Assume layer $\ell$ is transmitted over a subchannel which adopts the MCS with index $m$. The probability $\mathrm{P}_{i,j}^{(u,\ell)}$ of moving from state $s_{i}^{(u,\ell)}$ to state $s_{j}^{(u,\ell)}$ is
\begin{equation}
\mathrm{P}_{i,j}^{(u,\ell)} \!=\! \left\{ 
  \begin{array}{l l}
    (1-\mathrm{P}_{\ell,k_\ell-i}) [1-p_u(m)] & \quad \text{if $i-j = 1$}\\
    \mathrm{P}_{\ell,k_\ell-i}[1-p_u(m)] + p_u(m) & \quad \text{if $i = j$}\\
    0 & \quad \text{otherwise.}\\
  \end{array} \right.\!\!\!\!\!\!\! \label{eq.TransProb}
\end{equation}
\end{lemma}
\begin{IEEEproof}
Since the user AMC is in state $s_{i}^{(u,\ell)}$, user $u$ has collected $k_\ell-i$ linearly independent coded packets, i.e., $\mathrm{rank}(\mathbf{C}_u) = k_\ell-i$. As a new coded packet associated with layer $\ell$ is transmitted, we have just two possibilities:
\begin{itemize}
\item The rank of $\mathbf{C}_u$ is increased to $k_\ell-i+1$ - The coded packet is successfully received with probability $1-p_u(m)$, and it is linearly independent of the previously received coded packets with probability \mbox{$(1-\mathrm{P}_{\ell,k_\ell-i})$}. This event occurs with a probability equal to \mbox{$(1-\mathrm{P}_{\ell,k_\ell-i}) [1-p_u(m)]$}.
\item The rank of $\mathbf{C}_u$ does not change - That may occur because the coded packet is not successfully received or because it is linearly dependent of the previously received coded packets. This event occurs with a probability equal to $\mathrm{P}_{\ell,k_\ell-i}[1-p_u(m)] + p_u(m)$.
\end{itemize}
\end{IEEEproof}
From~\eqref{eq.TransProb}, we also understand that the probability of moving from state $s_{0}^{(u,\ell)}$ to another state is zero. Hence, $s_{0}^{(u,\ell)}$ represents the so-called \emph{absorbing} state of the AMC. All the remaining states $s_{1}^{(u,\ell)}, \ldots, s_{k_\ell}^{(u,\ell)}$ are commonly referred to as \emph{transient} states~\cite{FMC60}. The state transition diagram of the resulting AMC can be represented as reported in Fig.~\ref{fig:MarkovChain}.

From Lemma~\ref{l.transition}, it directly follows that the \mbox{$(k_\ell+1) \times (k_\ell+1)$} transition matrix $\mathbf{T}^{(u,\ell)}$ describing the AMC of user $u$ and associated with layer $\ell$ has the following structure in its \emph{canonical form}~\cite{FMC60}:
\setlength{\arraycolsep}{0.5em}\begin{equation}
\mathbf{T}^{(u,\ell)} \doteq
\left[
\begin{array}{c|c}
 1 & \mathbf{0}\\
 \hline
 \mathbf{R}^{(u,\ell)} & \mathbf{Q}^{(u,\ell)}
\end{array}
\right]\text{,}
\label{eq.P}
\end{equation}
where $\mathbf{Q}^{(u,\ell)}$ is the $k_\ell \times k_\ell$ transition matrix modeling the AMC process as long as it involves only transient states. The term $\mathbf{R}^{(u,\ell)}$ is a column vector of $k_\ell$ elements which lists all the probabilities of moving from a transient to the absorbing state. From~\cite[Theorem 3.2.4]{FMC60}, let define matrix $\mathbf{N}^{(u,\ell)}$ as
\begin{equation}
\mathbf{N}^{(u,\ell)} = \sum_{t=0}^\infty \left(\mathbf{Q}^{(u,\ell)}\right)^t = \Big[\mathbf{I} - \mathbf{Q}^{(u,\ell)}\Big]^{-1}. \label{eq.fundMatrix}
\end{equation}

Element $\mathrm{N}^{(u,\ell)}_{i,j}$ at the location $(i,j)$ of matrix $\mathbf{N}^{(u,\ell)}$ defines the average number of coded packet transmissions required for the process transition from state $s^{(u,\ell)}_{i}$ to state $s^{(u,\ell)}_{j}$, where both $s^{(u,\ell)}_{i}$ and $s^{(u,\ell)}_{j}$ are transient states. In particular, from Lemma~\ref{l.transition}, the following theorem holds
\begin{theorem}[{\hspace*{-1.3mm}\cite[Theorem~3.3.5]{FMC60}}]
If the AMC is in the transient state $s^{(u,\ell)}_{i}$, the average number of coded packet transmissions needed to get to state $s^{(u,\ell)}_{0}$ is
\begin{equation}
\tau^{(u,\ell)}_{i} = \left\{ 
  \begin{array}{l l}
  	0 & \quad \text{if $i = 0$}\\
    \displaystyle\sum_{j=1}^{i} \mathrm{N}^{(u,\ell)}_{i,j} & \quad \text{if $i = 1, \ldots, k_\ell$.}\\
  \end{array} \right.\label{eq.fundMatrix_1}
\end{equation}\label{th.335}
\end{theorem}
From~\eqref{eq.fundMatrix_1} and Theorem~\ref{th.335}, we prove the following corollaries.
\begin{corollary}\label{c.SNC}
In the case of S-RLNC, the average number $\tau^{(u,\ell)}_{\text{S-RLNC}}$ of coded packets transmissions needed by user $u$ to recover the source message layer $\ell$ is $\tau^{(u,\ell)}_{\text{S-RLNC}} = \tau^{(u,\ell)}_{k_\ell}$.
\end{corollary}
\begin{IEEEproof}
When the source node transmits the very first coded packet, user $u$ is in state $s^{(u,\ell)}_{k_\ell}$. That follows from the fact that the source node has not previously transmitted any coded packets, and, hence, $\mathrm{rank}(\mathbf{C}_u)$ is always equal to $0$.
\end{IEEEproof}

We remark that, in the case of S-SRLNC transmission, at the end of the systematic phase, user $u$ may have collected one or more source packets, implying that $\mathrm{def}(\mathbf{C}_u)$ may be smaller than $k_\ell$. In particular, if $\mathrm{def}(\mathbf{C}_u) < k_\ell$, the AMC will start from any of the states $s_{0}^{(u,\ell)}, \ldots, s_{k_\ell-1}^{(u,\ell)}$.
\begin{corollary}\label{c.sist}
Consider S-SRLNC. If systematic and non-systematic coded packets associated with source message $\ell$ are transmitted by means of the MCS with index $m$, the considered average number $\tau_{\text{S-SRLNC}}^{(u,\ell)}$ of systematic and coded packet transmissions needed to recover layer $\ell$ is
\begin{equation}
\tau_{\text{S-SRLNC}}^{(u,\ell)} = \sum_{i=0}^{k_\ell} \, \pi^{(u,\ell)}_{i} \, \left(k_\ell-i + \tau^{(u,\ell)}_{i}\right)
\label{eq.prob_ent_i2}
\end{equation}
where $\pi^{(u,\ell)}_{i}$ is the probability that the process associated with user $u$ starts from state $s^{(u,\ell)}_{i}$, given by
\begin{equation}
\!\pi^{(u,\ell)}_{i} = \binom{k_\ell}{i} p_u(m)^i \left[1-p_u(m)\right]^{k_\ell-i}, \quad \text{$i = 0, \ldots, k_\ell$}.\!\!\!\!\! \label{eq.prob_ent_i1}
\end{equation}
\end{corollary}
\begin{IEEEproof}
Assume that $u$ collects $k_\ell - i$ out of $k_\ell$ systematic packets. Hence, matrix $\mathbf{C}_u$ consists of $k_\ell - i$ linearly independent rows and, hence, the user AMC is in state $s^{(u,\ell)}_{i}$. In that case, from~\eqref{eq.fundMatrix_1}, we have that layer $\ell$ is recovered, on average, after $k_\ell - i+ \tau_i^{(u,\ell)}$ packet transmissions, namely, $k_\ell - i$ systematic packets plus $\tau_i^{(u,\ell)}$ coded packets. At the end of the systematic packet transmission phase, the AMC is in state $s^{(u,\ell)}_{i}$ with probability $\binom{k_\ell}{i}p_u(m)^i \left[1-p_u(m)\right]^{k_\ell-i}$, for $i = 0, \ldots, k_\ell$. Hence, the value of $\tau_{\text{S-SRLNC}}^{(u,\ell)}$ is obtained by simply averaging $k_\ell - i + \tau_i^{(u,\ell)}$ with the appropriate probability value of $\pi^{(u,\ell)}_{i}$, for $i = 0, \ldots, k_\ell$, as provided in~\eqref{eq.prob_ent_i2}. 
\end{IEEEproof}

\section{Sparse RLNC Optimization: Motivations and Resource Allocation Models}\label{sec.Opt}
Among the most effective ways of decreasing the computational complexity of the RLNC decoding operations, we consider the reduction of the number of source packets, and the increase of the sparsity of the non-degenerate coding vectors per source message layer. As discussed in Section~\ref{sec.SM}, we remark that as the MCS index $m_\ell$ used to transmit layer $\ell$ increases, the number $r(m_\ell)$ of useful bits carried by one resource block or, equivalently, forming a coded packet, is likely to increase. Given that coded and source packets have the same bit size, the value of $k_\ell$ is likely to decrease when $m_\ell$ increases. However, as $m_\ell$ increases, user PER related to the reception of subchannel $\ell$ is likely to increase, i.e., the fraction of multicast users regarding the reception of subchannel $\ell$ as acceptable is likely to decrease.

\begin{figure}[tb]
\centering
\subfloat[No. of coded packet transmissions]{\label{fig.motiv.tau}
\hspace{-1mm}\includegraphics[width=0.499\columnwidth]{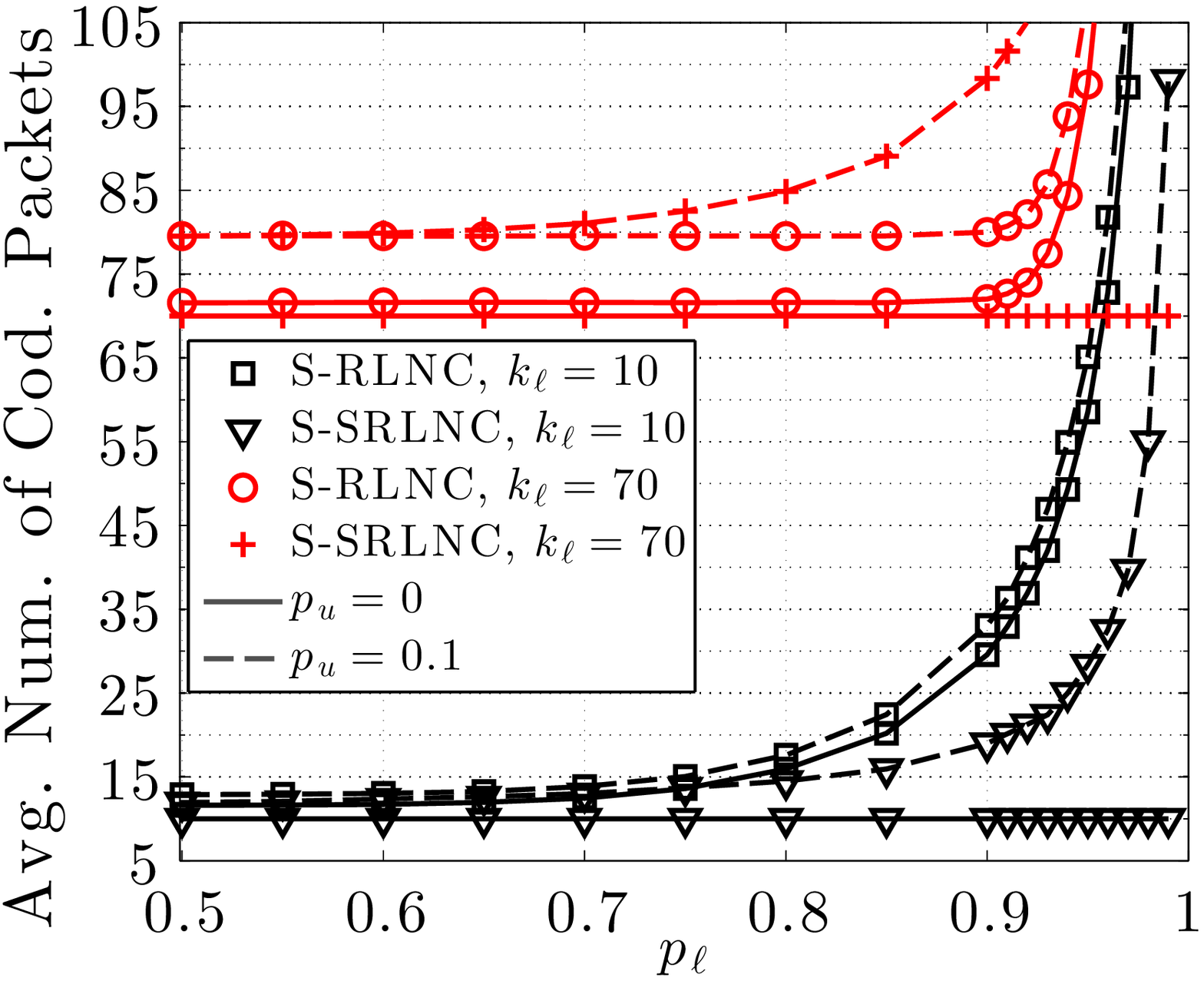}
}
\subfloat[No. of decoding operations]{\label{fig.motiv.ops}
\hspace{-2.1mm}\includegraphics[width=0.499\columnwidth]{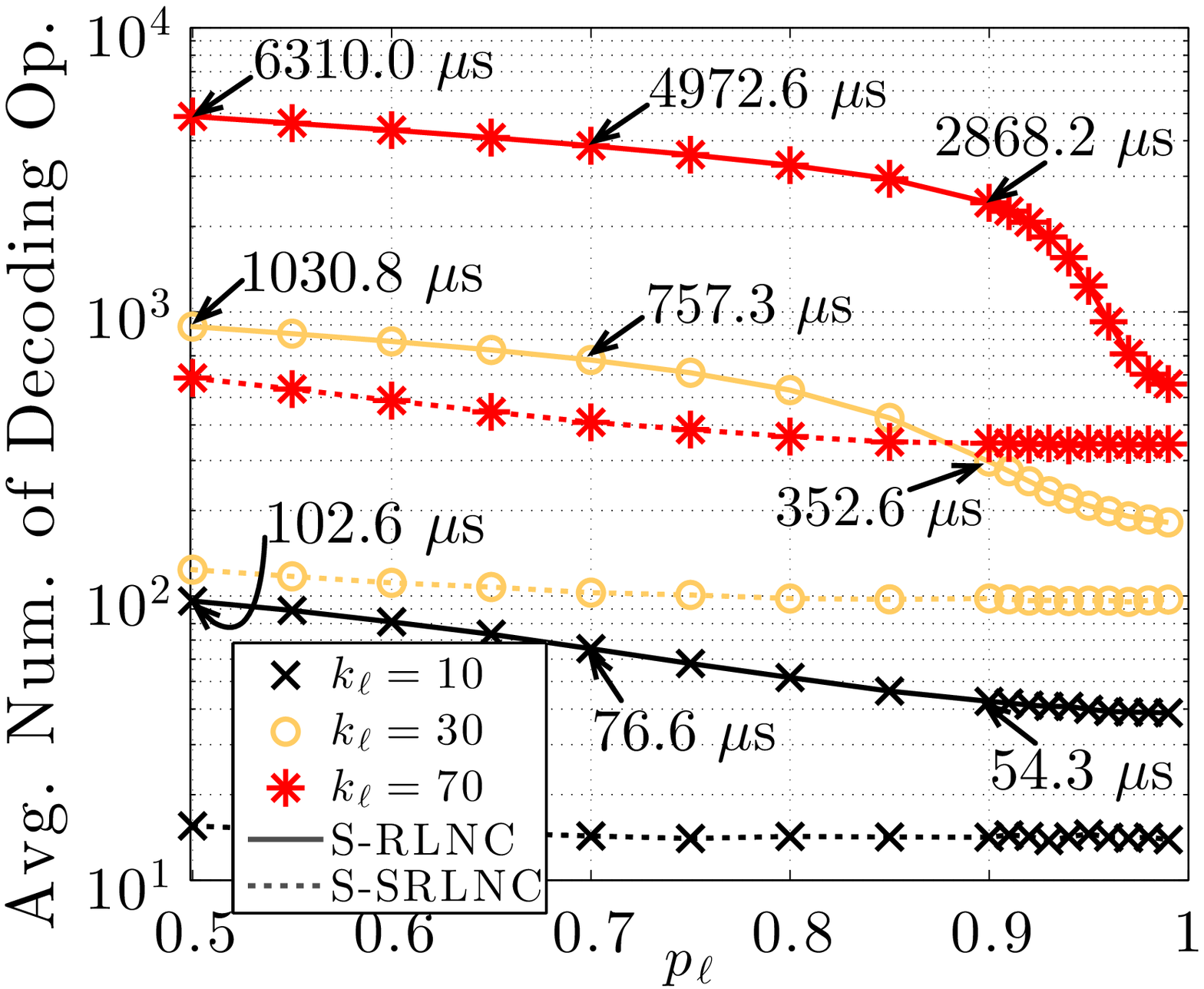}
}
\caption{Average number of coded packet transmissions and decoding operations, for $q = 2$. With regards the S-SRLNC scheme, the average number of decoding operations have been obtained by considering $p_u = 0.1$.}
\label{fig.motiv}
\end{figure}

It is worth noting that both the value of $k_\ell$ and the probability $p_\ell$ of selecting a coding coefficient equal to zero determine the average number of coded packet transmissions and the average number of decoding operations needed to recover layer $\ell$. With regards to the first aspect, Fig.~\ref{fig.motiv.tau} shows the value of $\tau^{(u,\ell)}_{\mathrm{S-RLNC}}$ and $\tau^{(u,\ell)}_{\mathrm{S-SRLNC}}$ as a function of $p_\ell$, for $q=2$, $k_\ell = \{10,70\}$ and a packet error probability $p_u = \{0, 0.1\}$, when \mbox{S-RLNC} or \mbox{S-SRLNC} is used. Curves have been obtained by computer simulations. More details about the simulation environment will be given in Section~\ref{sec.AR}. 

In the case of S-SRLNC, as discussed in Section~\ref{subsec.NCbac}, coded packets are transmitted after the systematic packets. Obviously, if $p_u = 0$, there is no need of transmitting coded packets as all the systematic packets are successfully received. That explains the reason way $\tau^{(u,\ell)}_{\mathrm{S-SRLNC}}$ is always equal to $k_\ell$, for $p_u = 0$. On the other hand, as the value of $p_u$ increases, the number of coded packets to be transmitted increases, as well. We also observe that, for the same value of $p_u$, $\tau^{(u,\ell)}_{\mathrm{S-SRLNC}}$ is smaller than or equal to $\tau^{(u,\ell)}_{\mathrm{S-RLNC}}$. That is given by the fact that, in the case of S-SRLNC, there is aways the possibility for a user to collect some systematic packets, which are obviously linearly independent.

Both with S-RLNC and S-SRLNC (for $p_u > 0$), we observe that if $p_\ell$ approaches $1$, then the average number of packet transmissions needed to recover layer $\ell$ increases. That is given by the fact that, coding vectors tend to be composed by all-zero. In addition, for a given value of $p_\ell$, as $k_\ell$ and/or $p_u$ decrease, the value of $\tau^{(u,\ell)}_{\mathrm{S-RLNC}}$ decreases. 

Fig.~\ref{fig.motiv.ops} shows the measured average number of decoding operations $\epsilon_{\mathrm{S-RLNC}}^{(\ell)}$ and $\epsilon_{\mathrm{S-SRLNC}}^{(\ell)}$ needed to recover layer $\ell$, in the S-RLNC and S-SRLNC case, respectively. Results are provided as a function of $p_\ell$, for $k_\ell = \{10, 30, 70\}$. Obviously, $\epsilon_{\mathrm{S-RLNC}}^{(\ell)}$ does not depend on the user PER but just on $k_\ell$ and $p_\ell$. In this paper, we will only refer to the fundamental finite field operations\footnote{Let $a, b, c$ be three elements in $\mathrm{GF}(q)$, we will consider the following operations: $a \cdot b$, $a + b$, $a - b$, $a + (b \cdot c)$ and $a - (b \cdot c)$.} performed by a network coding decoder based on the Gaussian Elimination principle, which represent the most computationally intensive part of the decoding process~\cite{KODO}. In particular, the more $p_\ell$ increases, the more the coding matrix $\mathbf{C}_u$ becomes sparser, and, consequently, the Gaussian Elimination requires a smaller number of iterations~\cite{LucaniNetCod}. That behavior is confirmed by Fig.~\ref{fig.motiv.ops}, $\epsilon_{\mathrm{S-RLNC}}^{(\ell)}$ decreases not only when $k_\ell$ decreases but also when $p_\ell$ increases.

In the case of S-SRLNC, the value of $\epsilon_{\mathrm{S-SRLNC}}^{(\ell)}$ is indeed affected by the user PER. The more $p_u$ increases, the more the number of successfully received systematic packets decreases and, the more the number of coded packets required to recover the layer increases. Hence, that corresponds to an increment in the value of $\epsilon_{\mathrm{S-SRLNC}}^{(\ell)}$. In particular, Fig.~\ref{fig.motiv.ops} shows the value of $\epsilon_{\mathrm{S-SRLNC}}^{(\ell)}$, for $p_u = 0.1$.

In the case of S-RLNC, in order to establish a link between the average number of decoding operations and the time needed to perform that number of decoding operations on a low-end device, Fig.~\ref{fig.motiv.ops} also reports the average processing time, for some $(p_\ell,k_\ell)$ pairs. We have referred to a Gaussian Elimination-based decoder run on a Raspberry Pi Model B~\cite{upton2014raspberry}. We note that there exists a linear relation between a reduction in the value of $\epsilon_{\mathrm{S-RLNC}}^{(\ell)}$ and in the average processing time. 

In the rest of the section, we will define a novel optimization model aiming to jointly optimize the sparsity of the code and the MCS index used to multicast each layer of the source message. The proposed model provides resource allocation solutions, which ensure that predetermined fractions of users recover sets of progressive layers, on average, within a given number of packet transmissions. In addition, the proposed model, at the same time, maximizes the sparsity and minimizes the total source message length.

\subsection{Proposed Resource Allocation Models}\label{subsec.alloc}
\begin{figure}[tb]
\centering
\includegraphics[width=1\columnwidth]{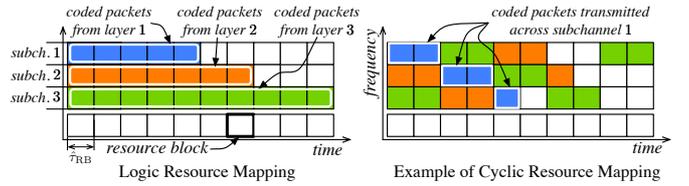}
\vspace{-3mm}\caption{Logic radio resource mapping (left-hand side) and an example of cyclic resource mapping (right-hand side), for $L = 3$.}
\label{fig.radioAlloc}
\end{figure}

From the \emph{logic perspective}, we refer to the radio resource mapping presented in Fig.~\ref{fig.radioAlloc} (left-hand side). As the resource block is our fundamental resource allocation unit, the time duration of each radio frame shall be an integer multiple of the resource block time duration $\Hat{\tau}_{\mathrm{RB}}$. Every $\Hat{\tau}_{\mathrm{RB}}$ seconds, the source mode transmits \emph{at most} one coded packet per-layer. We remark that the transmission of a message layer continues until the desired fraction of multicast users has recovered it (Section~\ref{subsec.NCbac}). As a result, the average number of packet transmissions can be easily related to the average time needed to recover a layer.

Even though all the resource blocks forming the same subchannel are mapped onto time contiguous OFDM symbols, they could span a different set of OFDM subcarriers every $\Hat{\tau}_{\mathrm{RB}}$ seconds. For instance, subchannels could cyclically span different frequency sub-bands, as shown in Fig.~\ref{fig.radioAlloc} (right-hand side). In that way, the transmission of the same subchannel across the same set of OFDM subcarriers is avoided. Hence, users experiencing poor channel conditions across specific OFDM subcarriers will not always be prevented from receiving the same message layer.

In order to optimize $m_\ell$ and, indirectly, $k_\ell$, the knowledge of the user propagation conditions is required. Obviously, the exact propagation conditions are unknown to the source node. However, modern communications standards allow users to periodically provide feedback about their average channel conditions across the whole transmission band\footnote{3GPP LTE and LTE-A standards refer to this kind of user channel feedback as wideband Channel Quality Indicators~\cite{sesia2011lte}.}. Generally, the PER experienced by $u$ is considered \emph{acceptable} if it is smaller than or equal to a threshold $\Hat{p}$. In the rest of the paper, we will refer to the principle adopted by the LTE-A standard, where any user $u$ provides as propagation condition feedback the greatest MCS index $\mathrm{M}_u$ such that $p_{u}(\mathrm{M}_u) \leq \Hat{p}$, defined as~\cite{sesia2011lte}:
\begin{equation}
\mathrm{M}_u \!=\! \left\{m \,| \,m \in [1, M] \wedge p_{u}(m) \leq \Hat{p}\wedge p_{u}(m+1) > \Hat{p}\right\}\!.\!\!\label{eq.cqi}
\end{equation}
For these reasons, if layer $\ell$ is transmitted with MCS index $m_\ell \leq \mathrm{M}_u$, $p_u(m_\ell)$ will be equal to or smaller than $\Hat{p}$. Given the ``aggregate nature'' of the user channel feedback, relation $p_{u}(\mathrm{M}_u) \leq \Hat{p}$ is to be considered valid across the whole system band. Hence, the notion of $\mathrm{M}_u$ is independent to the way subchannels are actually transmitted across each frame.

Owing to the lack of knowledge of the user PER, during the resource allocation phase, the source node approximates the user PER as
\begin{equation}
p_{u}(m_\ell) \cong \left\{ 
  \begin{array}{l l}
    \Hat{p} & \quad \text{if $m_\ell \leq \mathrm{M}_u$}\\
    1 & \quad \text{otherwise.}\\
  \end{array} \right. \label{eq.perOpt}
\end{equation}

In the case of S-RLNC, the proposed Sparsity-Tuning (ST) resource allocation model is defined as follows:
\begin{align}
	\text{ST} &  \quad  \mathop{\max_{p_1, \ldots, p_L}}_{m_1, \ldots, m_L} \,\,  \left\lVert\mathbf{p}\right\rVert_1 &\label{ST.of}\\
    \!\!\!\!\text{s.t.} & \,\,\sum_{u = 1}^U \delta\left(\sum_{t = 1}^\ell\tau_{\text{S-RLNC}}^{(u,t)} \leq \sum_{t=1}^\ell\Hat{\tau}_t\right) \geq \sum_{t=1}^\ell\Hat{U}_t,  \quad\text{$\ell = 1, \ldots, L$}\label{ST.c1}\\
                      &   \,\,q^{-1} \leq p_\ell < 1 \hspace{35mm} \text{$\ell = 1, \ldots, L$}\label{ST.c2}\\
                      &   \,\,m_\ell \in \{1, \ldots, M\} \hspace{29mm} \text{$\ell = 1, \ldots, L$}\label{ST.c3}
\end{align}
where objective function~\eqref{ST.of} maximizes the $1$-norm of vector $\mathbf{p} = \{p_1, \ldots, p_L\}$, which can be equivalently expressed as $\sum_{\ell = 1}^L p_\ell$. Term $\delta(t)$ is an indication function that is equal to $1$ if statement $t$ is true, otherwise it is equal to $0$. Parameters $\Hat{\tau}_\ell$ and $\Hat{U}_\ell$ represent the maximum number of coded packet transmissions needed to recover (on average) message layer $\ell$ and the minimum number of users that shall recover layer $\ell$, respectively. For these reasons, the left-hand side of constraint~\eqref{ST.c1} represents the number of multicast users that can recover layers $1, \ldots, \ell$, on average, in at most $\sum_{t=1}^\ell\Hat{\tau}_t$ coded packet transmissions. As a result, constraint~\eqref{ST.c1} ensures that the number of multicast users achieve \emph{QoS level} $\ell$ is at least equal to $\sum_{t=0}^\ell\Hat{U}_t$. Since user $u$ can only achieve QoS level $\ell$ if all the layers $1, \ldots, \ell$ have been recovered, it would be pointless to recover layer $\ell$ before layer $\ell - 1$. For the same reasons, there is no point in having situations where the fraction of users recovering layer $\ell$ is greater than the fraction of users recovering $\ell - 1$. Hence, it is reasonable to assume that the relations $\Hat{U}_{\ell-1} \geq \Hat{U}_{\ell}$ and $m_{\ell-1} \leq m_{\ell}$ hold, for $\ell = 2, \ldots, L$.
Furthermore, constraint~\eqref{ST.c2} avoids both dense coding vectors (i.e., $p_\ell < q^{-1}$) and all-zero coding vectors (i.e., $p_\ell=1$). Then constraint~\eqref{ST.c3} remarks that variable $m_\ell$ can only take values in range $1, \ldots, M$. The ST problem can also be defined for the case of S-SRLNC by simply replacing in constraint~\eqref{ST.c1} the term $\tau_{\text{S-RLNC}}^{(u,t)}$ with $\tau_{\text{S-SRLNC}}^{(u,t)}$. We observe that the selection of parameters $\Hat{\tau}_\ell$ and $\Hat{U}_\ell$, for $\ell = 1, \ldots, L$, allow the ultra-reliable service to be delivered, by meeting the Service Level Agreements (SLAs) between the service provider and the users. In our case, SLAs imposes the minimum fraction of users that shall achieve target QoS levels and the maximum time needed (on average) to do so.

Because of constraint~\eqref{ST.c1}, the ST problem presents vast coupling constraints among the whole set of optimization variables. In spite of the apparent optimization complexity, we will show that the ST problem can be efficiently solved, both in the case of S-RLNC and S-SRLNC, by decomposing it into subproblems of a reduced complexity. In order to do so, it is worth solving the Layer Sparsity Maximization (LSM) problem associated with user $u$, MCS index $m$ and layer $\ell$. We will eventually refer to the LSM problem to solve the SM problem. In particular, the LSM problem is defined as follows:
\begin{align}
	\text{LSM-$(\ell,u,m)$} &  \quad  \max_{p_\ell} \,\,  p_\ell \label{LSM.of}\\
    \text{s.t.} &   \quad \tau_{\text{S-RLNC}}^{(u,\ell)} \leq \Hat{\tau}_\ell\label{LSM.c1}\\
                      &   \quad q^{-1} \leq p_\ell \leq 1
\end{align}
From Corollary~\ref{c.SNC}, we have that $\tau_{\text{S-RLNC}}^{(u,\ell)}$ is defined as a sum of terms from matrix $\mathbf{N}^{(u,\ell)}$. In the following, we equivalently rewrite constraint~\eqref{LSM.c1} in order to avoid the explicit inversion of $\mathbf{I} - \mathbf{Q}^{(u,\ell)}$ in~\eqref{eq.fundMatrix}, and we prove the convexity of LSM-$(\ell,u,m)$.

We define the $k_\ell \times k_\ell$ matrix $\mathbf{W}^{(u,\ell)}$ as $\mathbf{W}^{(u,\ell)} = \mathbf{I} - \mathbf{Q}^{(u,\ell)}$. From~\eqref{eq.fundMatrix}, we have that $\mathbf{N}^{(u,\ell)} = (\mathbf{W}^{(u,\ell)})^{-1}$. 
Let $\mathrm{Q}^{(u,\ell)}_{i,j}$ be the $(i,j)$-th element of matrix $\mathbf{Q}^{(u,\ell)}$. From~\eqref{eq.TransProb} and~\eqref{eq.P}, we have that $\mathbf{Q}^{(u,\ell)}$ is a non-negative lower-triangular matrix with the following structure:
\setlength{\arraycolsep}{0.12em}\begin{equation}
{\!\!\!\!\!\!\!\!\!\mathbf{Q}^{(u,\ell)} \!=\!\begin{bmatrix}
  \mathrm{Q}^{(u,\ell)}_{1,1} & 0                           & \cdots & 0 & 0\\
  \mathrm{Q}^{(u,\ell)}_{2,1} & \mathrm{Q}^{(u,\ell)}_{2,2}                  & \cdots & 0 & 0\\
  \vdots    & \vdots            & \ddots & \vdots & \vdots\\
  0         & 0         & 0        & \mathrm{Q}^{(u,\ell)}_{k_\ell,k_\ell-1} & \mathrm{Q}^{(u,\ell)}_{k_\ell,k_\ell}
 \end{bmatrix}\!\!.\!\!\!\!\!\!\!\!}\label{eq.Q_2}
\end{equation}
Hence, for $i$ and $j = 1, \ldots, k_\ell$, element $(i,j)$ of $\mathbf{W}^{(u,\ell)}$ is
\setlength{\arraycolsep}{0.11em}\begin{equation}
  \!\mathrm{W}^{(u,\ell)}_{i,j} =\! \left\{ 
  \begin{array}{l l}
    -(1 - p_u(m)) (1-\mathrm{P}_{\ell,k_\ell-i}) & \quad \text{if $i-j = 1$}\\
    (1 - p_u(m)) (1-\mathrm{P}_{\ell,k_\ell-i}) & \quad \text{if $i = j$}\\
    0 & \quad \text{otherwise.}
  \end{array} \right.\!\label{eq.W}
\end{equation}
From~\eqref{eq.fundMatrix}, the following relation holds:
\begin{equation}
\mathbf{W}^{(u,\ell)} \cdot \mathbf{N}^{(u,\ell)} = \mathbf{I}. \label{eq.sistems}
\end{equation}
Relation~\eqref{eq.sistems} defines a set of $k_\ell$ disjoint parametric systems of equations, where $p_\ell$ is the system parameter and the elements of $\mathbf{N}^{(u,\ell)}$ are the system unknowns. System $s$, for \mbox{$s = 1, \ldots, k_\ell$}, consists of $k_\ell - s + 1$ equations. In particular, the $i$-th equation of system $s$, for $i = s, \ldots, k_\ell$, is defined as:
\begin{equation}
\displaystyle\sum_{j = s}^{i} \mathrm{W}^{(u,\ell)}_{i,j} \, \mathrm{N}^{(u,\ell)}_{j,s} = \delta(i = s).\label{eq.systems_1}
\end{equation}
From~\eqref{eq.Q_2},~\eqref{eq.W}, the solution of system $s$ can be expressed as
\begin{equation}\label{eq.Nsol}
\!\mathrm{N}^{(u,\ell)}_{i,s} \!=\! \left\{ 
  \begin{array}{l l}
    \displaystyle\left[(1-p_u(m)) (1-\mathrm{P}_{\ell,k_\ell-s})\right]^{-1} & \text{if $i = 1, \ldots, k_\ell$}\\
    0 & \text{otherwise.}\\
  \end{array} \right.\!\!\!\!\!\!
\end{equation}
As a result we can prove the following lemma.

\begin{lemma}\label{lemma.cvx}
The LSM-$(\ell,u,m)$ problem is convex. In addition, the optimum solution of the problem is the real root of 
\begin{equation}
\sum_{i = 0}^{k_\ell} \left[(1-p_u(m)) (1-\mathrm{P}_{\ell, k_\ell-i})\right]^{-1} - \Hat{\tau} _\ell= 0,\label{eq.solCvx}
\end{equation}
which is greater than or equal to $q^{-1}$ and smaller than $1$.
\end{lemma}
\begin{IEEEproof}
From Corollary~\ref{c.SNC} and~\eqref{eq.Nsol}, we have that $\tau_{\text{S-RLNC}}^{(u,\ell)}$ can be equivalently rewritten as
\begin{equation}
\tau_{\text{S-RLNC}}^{(u,\ell)} = \displaystyle\sum_{i = 0}^{k_\ell} \left[(1-p_u(m)) (1-\mathrm{P}_{\ell,k_\ell-i})\right]^{-1}.\label{eq.sNCTau}
\end{equation}
Since we refer to the approximation as in~\eqref{eq.LBapp}, $\mathrm{P}_{\ell,k_\ell-i}$ is the non-negative power of a pointwise maximization of two convex functions. Hence, $\mathrm{P}_{\ell,k_\ell-i}$ is convex with respect to $p_\ell$. Consider function $(1-p_u(m)) (1-\mathrm{P}_{\ell,k_\ell-i})$ of~\eqref{eq.sNCTau}. Since $\mathrm{P}_{\ell,k_\ell-i}$ is convex, function $(1-p_u(m)) (1-\mathrm{P}_{\ell,k_\ell-i})$ is concave and, hence, $\left[(1-p_u(m)) (1-\mathrm{P}_{\ell,k_\ell-i})\right]^{-1}$ is convex. As a result, $\tau_{\text{S-RLNC}}^{(u,\ell)}$, expressed as in~\eqref{eq.sNCTau}, is a non-negative weighted sum of convex functions, which is a convex function. For these reasons, it follows that the LSM-$(\ell,u,m)$ problem is convex~\cite{Boyd:2004:CO:993483}. From~\eqref{eq.sNCTau}, we rewrite constraint~\eqref{LSM.c1} as $\sum_{i = 1}^{k_\ell} \left[(1-p_u(m)) (1-\mathrm{P}_{\ell,k_\ell-i})\right]^{-1} \leq \Hat{\tau}_\ell$. Because of the convexity of LSM-$(\ell,u,m)$, we have that the optimum solution of the problem is given by the real root of~\eqref{eq.solCvx}, which belongs to $[q^{-1}, 1)$.
\end{IEEEproof}

The LSM-$(\ell,u,m)$ problem can be adapted to the S-SRLNC case by simply replacing constraint~\eqref{LSM.c1} with $\tau_{\text{S-SRLNC}}^{(u,\ell)} \leq \Hat{\tau}_\ell$. The resulting optimization problem can be solved as follows.
\begin{lemma}\label{lemma.cvxsys}
In the S-SRLNC case, the resulting LSM-$(\ell,u,m)$ problem is convex, and its optimal solution is the real root, greater than or equal to $q^{-1}$ and smaller than $1$, of the following equation:
\setlength{\arraycolsep}{0.0em}
\begin{eqnarray}
\hspace*{-1mm}\displaystyle\sum_{i=0}^{k_\ell} \, \pi^{(u,\ell)}_{i} \, (k_\ell - i) + \hspace*{51mm}\notag\\
\hspace*{-1mm}+\sum_{i = 1}^{k_\ell} \pi^{(u,\ell)}_{i} \displaystyle\sum_{j = 1}^{i} \left[(1-p_u(m)) (1-\mathrm{P}_{\ell,k_\ell-j})\right]^{-1} - \Hat{\tau}_\ell = 0.\hspace{4mm}\label{eq.solCvxsys}
\end{eqnarray}
\end{lemma}
\begin{IEEEproof}
From Corollary~\ref{c.sist} and~\eqref{eq.Nsol}, $\tau_{\text{S-SRLNC}}^{(u,\ell)}$ can be expressed as
\begin{multline}
\hspace{0.1mm}\tau_{\text{S-SRLNC}}^{(u,\ell)} = \pi^{(u,\ell)}_{0} \, k_\ell +\sum_{i = 1}^{k_\ell} \Bigg\{\pi^{(u,\ell)}_{i} (k_\ell - i) \,+\\
\hspace{5mm}+\, \pi^{(u,\ell)}_{i}\sum_{j = 1}^{i} \left[(1-p_u(m)) (1-\mathrm{P}_{\ell,k_\ell-j})\right]^{-1}\Bigg\}.\!\!\!\label{eq.systNCTau}
\end{multline}
Likewise Lemma~\ref{lemma.cvx}, $\tau_{\text{S-SRLNC}}^{(u,\ell)}$ is convex because it is the non-negative weighted sum of convex functions. Then the proof follows exactly the same reasoning as in the proof of Lemma~\ref{lemma.cvx}.
\end{IEEEproof}

Once more, consider the ST problem and the following remark.
\begin{lemma}\label{lem.3}
Constraint~\eqref{ST.c1} of the ST problem can be equivalently rewritten as
\begin{equation}
\sum_{u = 1}^U \delta\left(\tau_{\text{S-RLNC}}^{(u,\ell)} \leq \Hat{\tau}_\ell\right) \geq \Hat{U}_\ell, \quad\quad \text{$\ell = 1, \ldots, L$.} \label{eq.newConst}
\end{equation}
or restated for the S-SRLNC case, in a similar way.
\end{lemma}
\begin{IEEEproof}
From Section~\ref{subsec.alloc}, relation $\tau_{\text{S-RLNC}}^{(u,t)} \leq \Hat{\tau}_t$ shall hold, for at least $\Hat{U}_t$ users. Hence, the complete statement of the argument of function $\delta(\cdot)$ in~\eqref{ST.c1} is equivalent to the following system of inequalities
\begin{equation}
\left\{ 
  \begin{array}{l l}
    \displaystyle\sum_{t = 1}^\ell\tau_{\text{S-RLNC}}^{(u,t)} \leq \sum_{t=1}^\ell\Hat{\tau}_t &\\
    \displaystyle\tau_{\text{S-RLNC}}^{(u,t)} \leq \Hat{\tau}_t, & \quad\text{for $t = 1, \ldots, \ell$.}\\
  \end{array} \right. \label{eq.TransProb}
\end{equation}
We observe that the first inequality is made redundant by the remaining ones. Hence,~\eqref{ST.c1} can be rewritten as 
\begin{equation}
\sum_{u = 1}^U \delta\left(\bigwedge_{t = 1}^\ell \tau_{\text{S-RLNC}}^{(u,t)} \leq \Hat{\tau}_t \right) \geq \sum_{t=1}^\ell\Hat{U}_t, \quad\text{for $\ell = 1, \ldots, L$}, \label{ST.c1.RW}
\end{equation}
where the leftmost term still counts exactly the same number of users achieving QoS level $\ell$ as in~\eqref{ST.c1}. Consider layer $t$, it shall be received by at least $\Hat{U}_t$ users, for $t = 1, \ldots, L$. Hence, the complete statement of~\eqref{ST.c1.RW}, for a given $\ell$, is
\begin{equation}
\left\{ 
  \begin{array}{l l}
    \displaystyle\sum_{u = 1}^U \delta\left(\bigwedge_{t = 1}^\ell \tau_{\text{S-RLNC}}^{(u,t)} \leq \Hat{\tau}_t \right) \geq \sum_{t=1}^\ell\Hat{U}_t, & \\
    \displaystyle\sum_{u = 1}^U \delta\left(\tau_{\text{S-RLNC}}^{(u,t)} \leq \Hat{\tau}_t\right) \geq \Hat{U}_t, & \quad\text{for $t = 1, \ldots, \ell$.}\\
  \end{array} \right. \label{eq.fp}
\end{equation}
We remark that relations $\Hat{U}_{\ell-1} \geq \Hat{U}_{\ell}$ and $m_{\ell-1} \leq m_{\ell}$ hold, for $\ell = 2, \ldots, L$. In addition, from the considered PER model~\eqref{eq.perOpt}, we have that the set of users achieving QoS level $\ell$ entirely contains those achieving QoS levels $1, \ldots, \ell - 1$. Hence, the first inequality of~\eqref{eq.fp} is made redundant by the following ones. That completes the proof. This proof can be similarly restated for the S-SRLNC case.
\end{IEEEproof}

From Lemma~\ref{lem.3}, ST can be decomposed into $L$ independent optimization problems ST-$(1)$, \ldots, ST-$(L)$, where the ST-$(\ell)$ problem: (i)  refers to the video layer $\ell$, (ii) has the goal of maximizing $p_\ell$, and (iii) refers to just the constraints of ST that are related to layer $\ell$. ST-$(\ell)$ problem can be solved as follows.
\begin{remark}\label{rem.2}
From Lemmas~\ref{lemma.cvx} and~\ref{lemma.cvxsys}, we have that $\tau_{\text{S-RLNC}}^{(u,\ell)}$ and $\tau_{\text{S-SRLNC}}^{(u,\ell)}$ are non-decreasing functions with respect to $p_\ell$, for $q^{-1} \leq p_\ell < 1$. In addition, for a given value of $p_\ell$, we remark that as $m_\ell$ increases, the value of $\tau_{\text{S-RLNC}}^{(u,\ell)}$ will decrease as well (Section~\ref{sec.SM}). Hence, ST-$(\ell)$ is solved by the pair $(m_\ell,p_\ell)$ characterized by the greatest values of $m_\ell$ and $p_\ell$ such that relations $\tau_{\text{S-RLNC}}^{(u,\ell)} \leq \Hat{\tau}_\ell$ or $\tau_{\text{S-SRLNC}}^{(u,\ell)} \leq \Hat{\tau}_\ell$ hold, for at least $\Hat{U}_\ell$ users. In particular, ST-($\ell$) can be solved by resorting to LSM problems as follows.
For any $m_\ell = 1, \ldots, M$ and $\ell = 1, \ldots, L$, let $\mathcal{U}_{m_{\ell}}$ signify the set of users such that $\mathrm{M}_u \geq m_\ell$.
\begin{enumerate}[1.]
\item Let us solve LSM-$(\ell,u,m)$, for a user $u \in \mathcal{U}_{m_\ell}$ and \mbox{$m = m_\ell$}. Let $p^*_{\ell,m_\ell}$ be the optimum solution of LSM-$(\ell,u,m_\ell)$. 
If S-RLNC is in use then the value of $p^*_{\ell,m_\ell}$ is derived as provided by Lemma~\ref{lemma.cvx}.
On the other hand, if S-SRLNC is in use then we will refer to Lemma~\ref{lemma.cvxsys}, for the computation of $p^*_{\ell,m_\ell}$.
Since $p_u(m)$ is approximated as in~\eqref{eq.perOpt}, the solution $p^*_{\ell,m_\ell}$ will always be the same, for every user in $\mathcal{U}_{m_\ell}$.
\item For any $m_\ell = 1, \ldots, M$ such that $|\mathcal{U}_{m_\ell}| \geq \Hat{U}_\ell$ and an optimum solution $p^*_{\ell,m_\ell}$ exists, the pair $(m_\ell,p^*_{\ell,m_\ell})$ is an optimum solution of ST-$(\ell)$. Among the optimum solutions of problem ST-$(\ell)$, we choose the pair $(m_\ell,p^*_{\ell,m_\ell})$ associated with the greatest MCS index, i.e., we consider the solution that ensures the smallest value of $k_\ell$ (see Section~\ref{sec.SM}).
\end{enumerate}
The process is repeated to solve any problem ST-$(\ell)$, for \mbox{$\ell = 1, \ldots, L$} and, hence, to solve problem ST. We observe that, for a given value of $m_\ell$, the pair $(m_\ell,p^*_{\ell,m_\ell})$ may not exist. That can happen because: (i) the value of $\Hat{\tau}_{\ell}$ is too small and the average number of coded packet transmissions always exceed $\Hat{\tau}_{\ell}$, for $q^{-1} \leq p_\ell < 1$, and/or (ii) the target user coverage $\Hat{U}_\ell$ is too big (constraint~\eqref{eq.newConst} is not met), given the overall user propagation conditions and, hence, the MCSs that can be used in a considered scenario.
\end{remark}

For these reasons and, in particular, from Lemmas~\ref{lemma.cvx},~\ref{lemma.cvxsys} and~\ref{lem.3}, it is immediate to prove the following theorem.
\begin{theorem}
Both in the S-RLNC and S-SRLNC cases, the resource allocation solution of ST problem derived by Remark~\ref{rem.2}, for any $\ell = 1, \ldots, L$,  is optimal and characterized by the greatest MCS indexes, i.e., the derived optimal solution ensures the smallest values of $k_\ell$, for $\ell = 1, \ldots, L$.
\end{theorem}

\section{Numerical Results}\label{sec.AR}
\subsection{Assessment of the Performance Model}\label{subsec.ass}
\begin{figure}[tb]
\centering

\subfloat[$q = 2$]{\label{fig.val.2}
	\includegraphics[width=0.95\columnwidth]{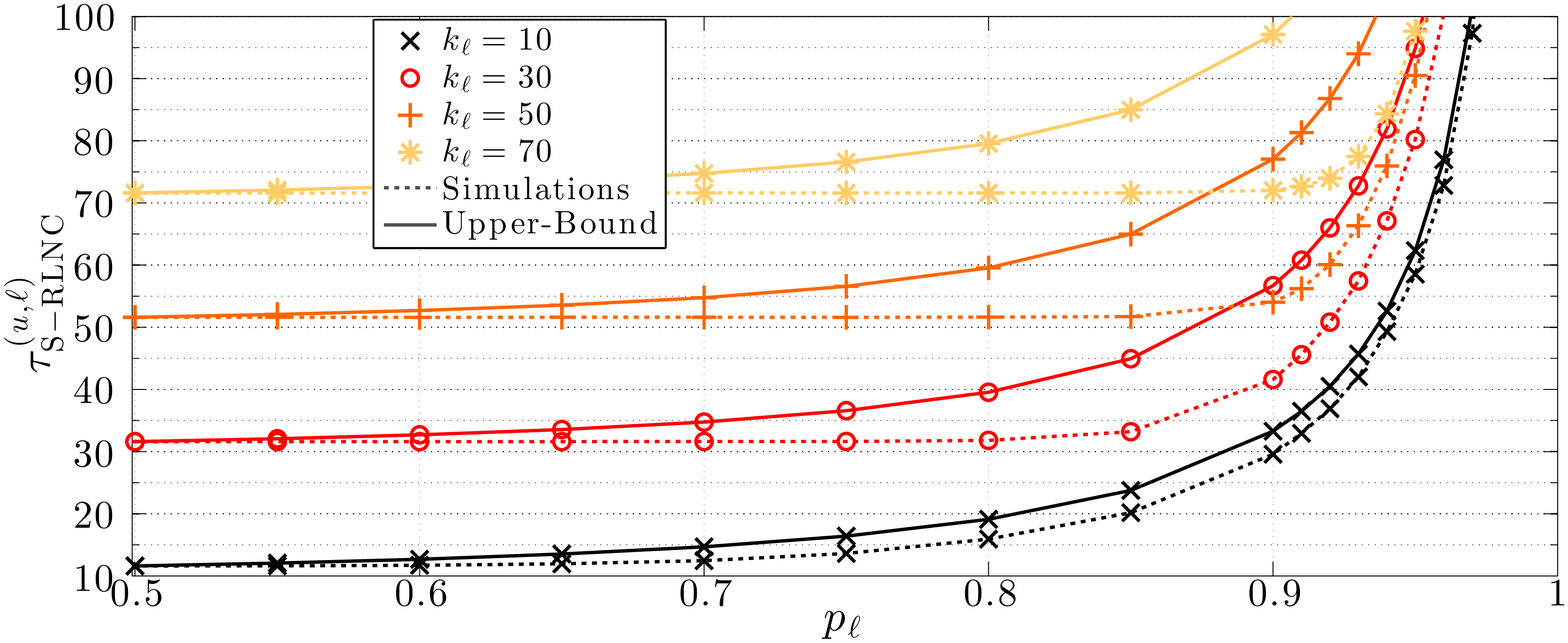}\vspace{-3mm}
}\\
\vspace{-3mm}\subfloat[$q = 2^8$]{\label{fig.val.256}
	\includegraphics[width=0.95\columnwidth]{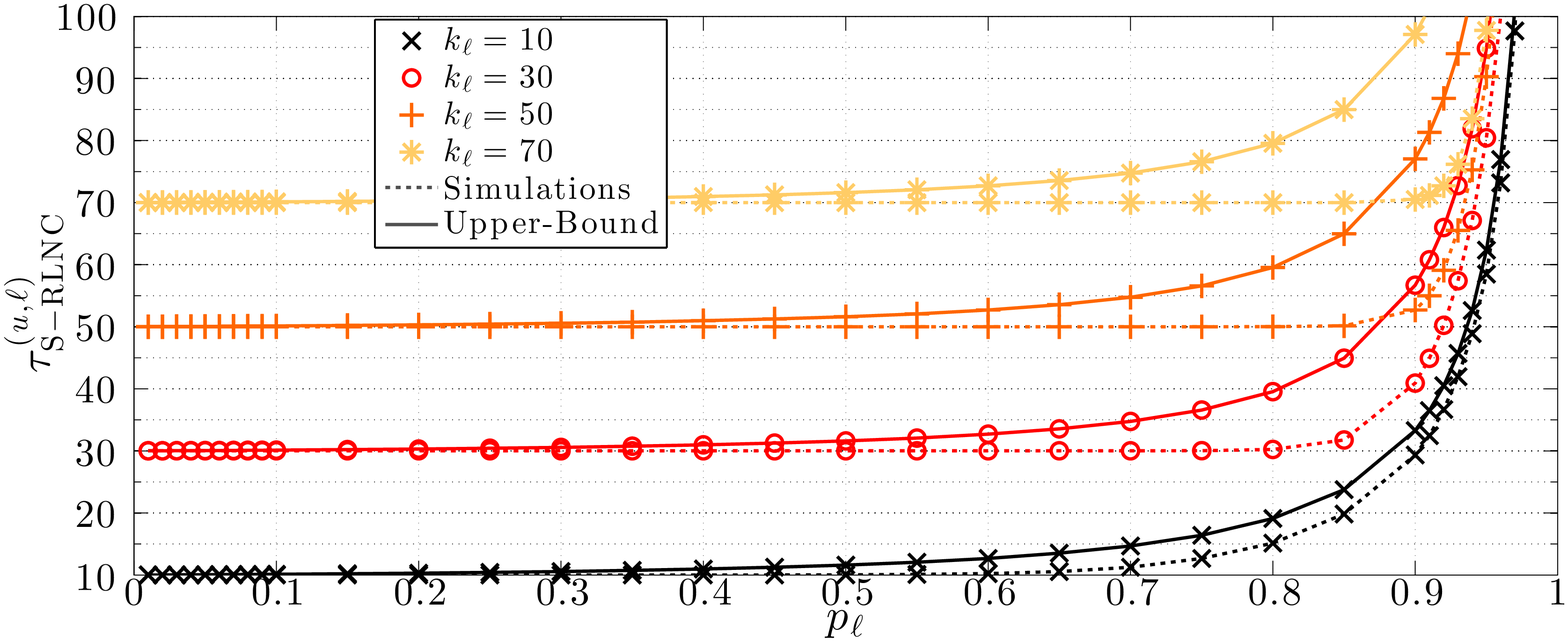}\vspace{-3mm}
}
\caption{Average number of coded packet transmissions vs. the average number of coded packet transmissions obtained by referring to the approximation as in~\eqref{eq.LBapp}, for $q = 2$ and $2^8$.}
\label{fig.val}
\end{figure}

We recall from Section~\ref{subsec.AMC} that we mitigated the lack of an accurate expression of the probability $\mathrm{P}_{\ell,t}$ that a sparse random $(t+1) \times k_\ell$ matrix is not full-rank over $\mathrm{GF}(q)$, given that the first $t$ rows are linearly independent. In particular, we upper-bounded the value of $\mathrm{P}_{\ell,t}$ by referring to the approximation in~\eqref{eq.LBapp}. Hence, the average user delay values $\tau^{(u,\ell)}_{\text{S-RLNC}}$ (Corollary~\ref{c.SNC}) and $\tau^{(u,\ell)}_{\text{S-SRLNC}}$ (Corollary~\ref{c.sist}) are expected to be greater than or equal to the correspondent average user delay values obtained via computer simulations. In this paper, all the computer simulations rely on the encoders and decoders provided by the Kodo C++ network coding library~\cite{KODO}.

Fig.~\ref{fig.val} refers to a scenario, where a source message of $k_\ell \in \{10, 30, 50, 70\}$ source packets is transmitted to a user by means of S-RLNC, over a fully reliable channel (i.e., the user PER is equal to $0$). In particular, Fig.~\ref{fig.val.2} compares, for $q = 2$, the value of $\tau^{(u,\ell)}_{\text{S-RLNC}}$ as in Corollary~\ref{c.SNC} with that obtained by simulations, as a function of the probability $p_\ell$ of selecting a zero coding coefficient. Fig.~\ref{fig.val.256} reports the same performance comparison, in the case of $q = 2^8$. Figs.~\ref{fig.val.2} and~\ref{fig.val.256} show that, for $p_\ell = q^{-1}$, simulation and our theoretical upper-bound of $\tau^{(u,\ell)}_{\text{S-RLNC}}$ overlap. In fact, from Remark~\ref{rem.1}, in that case,~\eqref{eq.LBapp} no longer is an approximation. However, the gap between the theoretical upper-bound and simulation results increases, as $p_\ell$ becomes larger than $q^{-1}$.

Let us focus on S-RLNC such that $p_\ell \geq q^{-1}$, regardless of the value of $q$, we observe that the performance gap between the theoretical upper-bound and simulation results mainly depends only on the value of $k_\ell$ and $p_\ell$. On the other hand, for large values of $k_\ell$ (such as, $k_\ell \geq 50$) and $p_\ell$ ($p_\ell \geq 0.93$), the value of the performance gap, normalized with respect to $k_\ell$, is almost constant and equal to $0.53$.
In other words, the impact of $q$ on the performance gap is not pivotal and, at the same time, it is mainly proportional to $k_\ell$. Given that the simulation results reported in Section~\ref{subsec.fw} refer to values of $k_\ell$ and $p_\ell$ in the aforementioned ranges, that gives a clear upper-bound of the impact of our approximation onto the displayed performance, on a layer-basis. Since S-RLNC can be considered a special case of S-SRLNC, the aforementioned considerations also apply to the systematic case.

We observe that the theoretical upper-bound is no more than $33.4$\% higher than simulation results, in the considered cases. As one of the key aspects in our optimization framework is the enforcement of service coverage constraints, the adoption of the approximation as in~\eqref{eq.LBapp} will indeed not violate those service constraints. Ideally, if the exact expression of $\tau_{\text{S-RLNC}}^{(u,\ell)}$ and $\tau_{\text{S-SRLNC}}^{(u,\ell)}$ were known, we would get ST solutions characterised by a greater level of sparsity. In addition, from an implementation perspective, it is not feasible to tabulate the exact values of $\tau_{\text{S-RLNC}}^{(u,\ell)}$ and $\tau_{\text{S-SRLNC}}^{(u,\ell)}$ as a function of $p_\ell$ and $k_\ell$. In particular, we remark that the value of $k_\ell$ is given by the layer bit length and the adopted MCS, which cannot be determined in advance.

\subsection{Performance Evaluation of the Proposed Resource Allocation Models}\label{subsec.fw}
The performance of the proposed resource allocation modeling has been investigated in an LTE-A scenario composed by $19$ base stations arranged in two concentric rings and centered on a \emph{target base station}. Each base station manages three hexagonal sectors per cell. In addition, for the physical layer parameters, we referred to the 3GPP's benchmark Case~$1$ scenario~\cite{TR_36_814}, where base stations are characterized by an inter-site distance of $500$ m. In order to meet the LTE-A physical layer constraints, each coded packet is mapped on resource blocks spanning a bandwidth of $540$ kHz and $12$ OFDM symbols (lasting for $\Hat{\tau}_{\mathrm{RB}} = 10$ ms). In accordance to a well documented best practice in the deployment of LTE-A networks~\cite{sesia2011lte}, the reception of a resource block is regarded as acceptable when $\Hat{p}$ is equal to $0.1$. The reader who may want to have more details about the simulator and the considered low-level transmission parameters, can refer to~\cite[Section V]{jsacTassi}. Due to space limitations, all those details have been omitted.

In our performance investigation, we referred to a network scenario where the target base station multicasts a layered video stream to a user MG, also known as Single Cell-eMBMS (SC-eMBMS) transmission mode. Furthermore, we considered a user distribution characterized by the maximum heterogeneity from the point of view of the channel conditions. In particular, we refer to a MG of $U = 80$ users that are regularly placed along the symmetry axis of one sector controlled by the target base station. The first user is $90$ m apart from the center of the cell, and the distance between two consecutive users is $2$ m.

In this section, we consider two different video sequences (Stream A and Stream B) of $30$~s, compressed according to the H.264/SVC standard~\cite{h264}:
\begin{itemize}
\item Stream A~\cite{StreamA} - is a $L = 3$ video trace characterized by $\{b_1, b_2, b_3\} = \{702, 4841, 20584\}$ KBytes per layer.
\item Stream B~\cite{StreamA} - is a $L = 4$ video trace such that $\{b_1, b_2, b_3, b_4\} = \{702, 2138, 6001, 19384\}$ KBytes per layer.
\end{itemize}
The video traces implement the coarse grain scalability principle, which is a form of spatial scalability such that the combination of consecutive layers enhances the frame resolution. In addition, both video traces belong to the database presented in~\cite{6025326}, and developed for network performance evaluation purposes. The video traces have a resolution of $352 \times 288$, a Group of Picture size of $16$ frames and a video frame rate of $30$ fps.

In our numerical results, a $30$ s video trace defines one layered source message. Each video layer has the same duration of the whole video trace. For simplicity, we impose that each video layer shall be recovered by the same average number $\Hat{\tau}$ of coded packet transmissions, i.e., $\Hat{\tau_1} = \Hat{\tau_2} = \ldots = \Hat{\tau_L} = \Hat{\tau}$. From constraint~\eqref{ST.c1}, we have that QoS level $\ell$ shall be achieved, on average, in $\sum_{t = 1}^{\ell}\Hat{\tau}_t = \ell  \Hat{\tau}$ coded packet transmissions. Since the time duration of each resource block is fixed, it is immediate to equivalently express $\Hat{\tau}$ in seconds, denoted as ``$\Hat{\tau}$ (sec.)''.

We compared the optimized version of S-RLNC and S-SRLNC (see Section~\ref{subsec.alloc}) against their non-sparse versions. In order to provide a fair comparison among the strategies, when either RLNC or SRLNC is used, the MCS indexes $m_1, \ldots, m_L$ associated to the transmission of each video layer are optimized such that the service constraints are met, for $p_\ell = q^{-1}$.

Let us define the average transmission footprint $\tau^{(u,1:\ell)}$ as
\begin{equation}
\tau^{(u,1:\ell)} = \begin{cases}
			\displaystyle\sum_{t=1}^{\ell} \tau_{\mathrm{S-RLNC}}^{(u,t)}\text{,} \quad &\text{for S-RLNC}\\
			\displaystyle\sum_{t=1}^{\ell} \tau_{\mathrm{S-SRLNC}}^{(u,t)}\text{,} \quad &\text{for S-SRLNC.}
		 \end{cases}\label{eq.tau}
\end{equation}
That definition can be easily extended to the non-sparse version of RLNC and SRLNC by considering a value of $p_\ell$ equal to $q^{-1}$. We remark that in the rest of this section, all the user performance investigation has been carried out via computer simulations. The approximated performance modeling of Section~\ref{subsec.AMC} is used only by the target base station during the resource allocation operations.

\begin{figure}[tb]
\centering
\subfloat[RLNC and SRLNC]{\label{fig.vl3.del.ns}
\includegraphics[width=0.95\columnwidth]{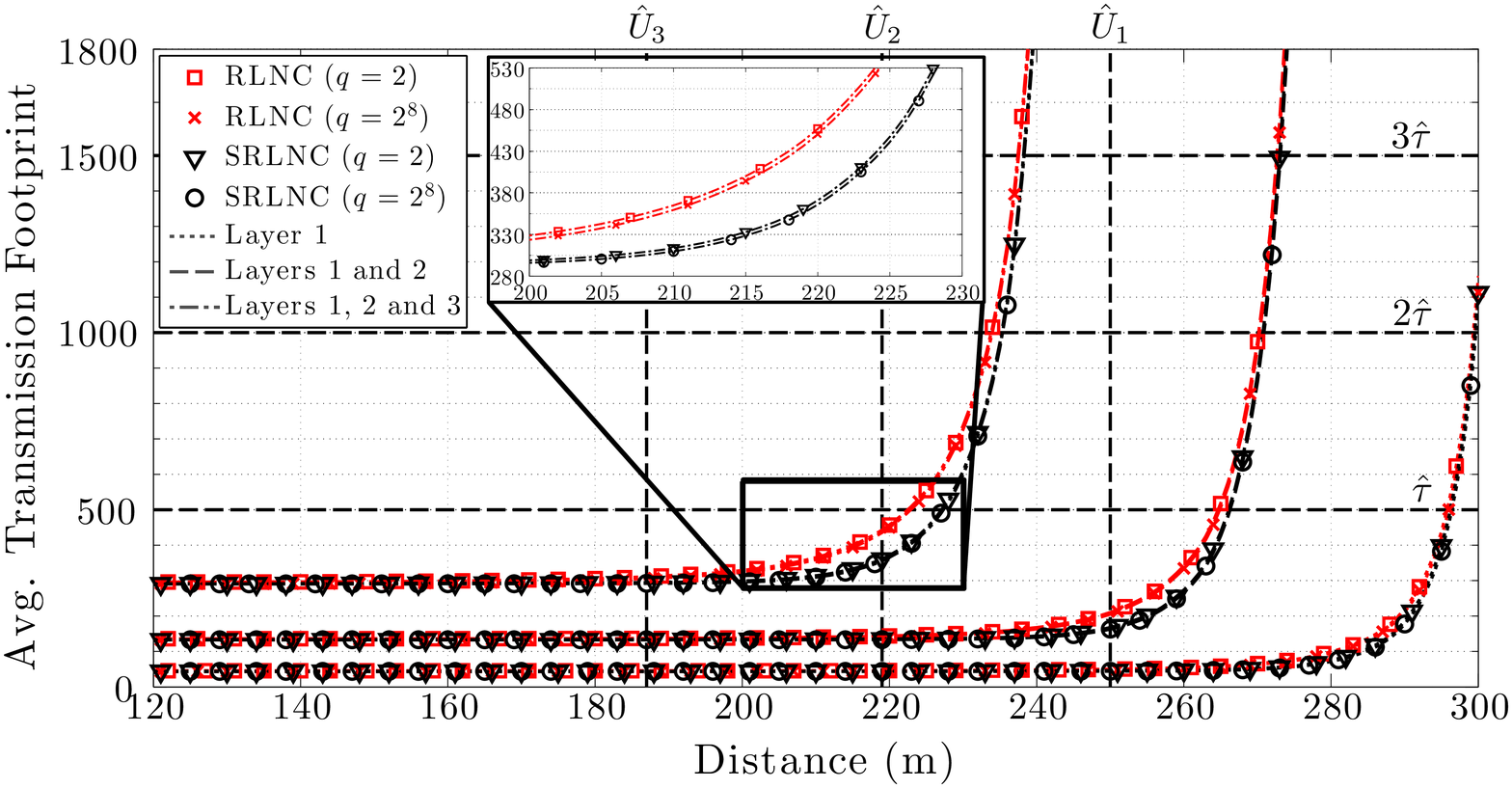}\vspace{-3mm}
}\\
\vspace{-3mm}\subfloat[S-RLNC and S-SRLNC]{\label{fig.vl3.del.s}
\includegraphics[width=0.95\columnwidth]{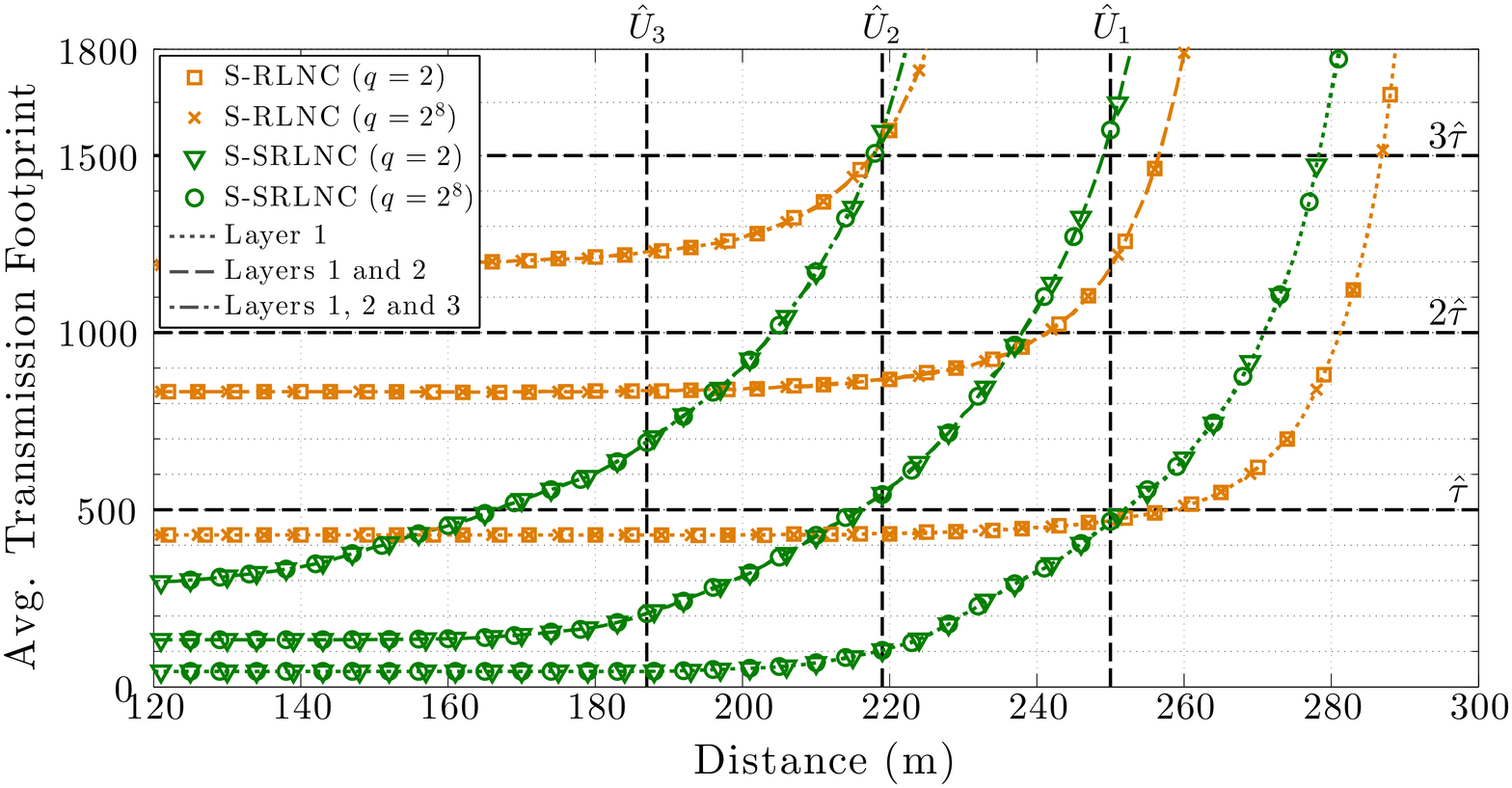}\vspace{-3mm}
}
\caption{Average transmission footprint $\tau^{(u,1:\ell)}$, expressed in terms of number of packet transmissions, in the case of Stream A, for $\ell = 1, \ldots, 3$, $q = 2$ and $2^8$.}
\label{fig.vl3.del}
\end{figure}

Figs.~\ref{fig.vl3.del.ns} and~\ref{fig.vl3.del.s} show the value of $\tau^{(u,1:\ell)}$ provided by all the considered network coding schemes when Stream A is multicast, in the case of $q = \{2, 2^8\}$ and for $\Hat{\tau}$ (sec.) equal to $0.5$ s. Since users are regularly distributed along with a segment connecting the target base station with the cell edge, we have that: (i) $\tau^{(u,1:\ell)}$ can be equivalently expressed as a function of the distance form the center of the cell, and (ii) the target number of users $\Hat{U}_\ell$ that shall receive video layer $\ell$ can be equivalently expressed in terms of distances from the center of the cell (vertical dashed lines in Figs.~\ref{fig.vl3.del.ns} and~\ref{fig.vl3.del.s}). On the other hand, the horizontal dashed lines in Figs.~\ref{fig.vl3.del.ns} and~\ref{fig.vl3.del.s} represents the maximum transmission footprint to achieve QoS level $\ell$, namely, $\sum_{t = 1}^{\ell}\Hat{\tau}_t$.

\begin{figure}[t]
\centering
\subfloat[{$\mathbb{E}[\mathrm{X}]$ values}]{\label{fig.r2.exp}
\hspace{-1mm}\includegraphics[width=0.499\columnwidth]{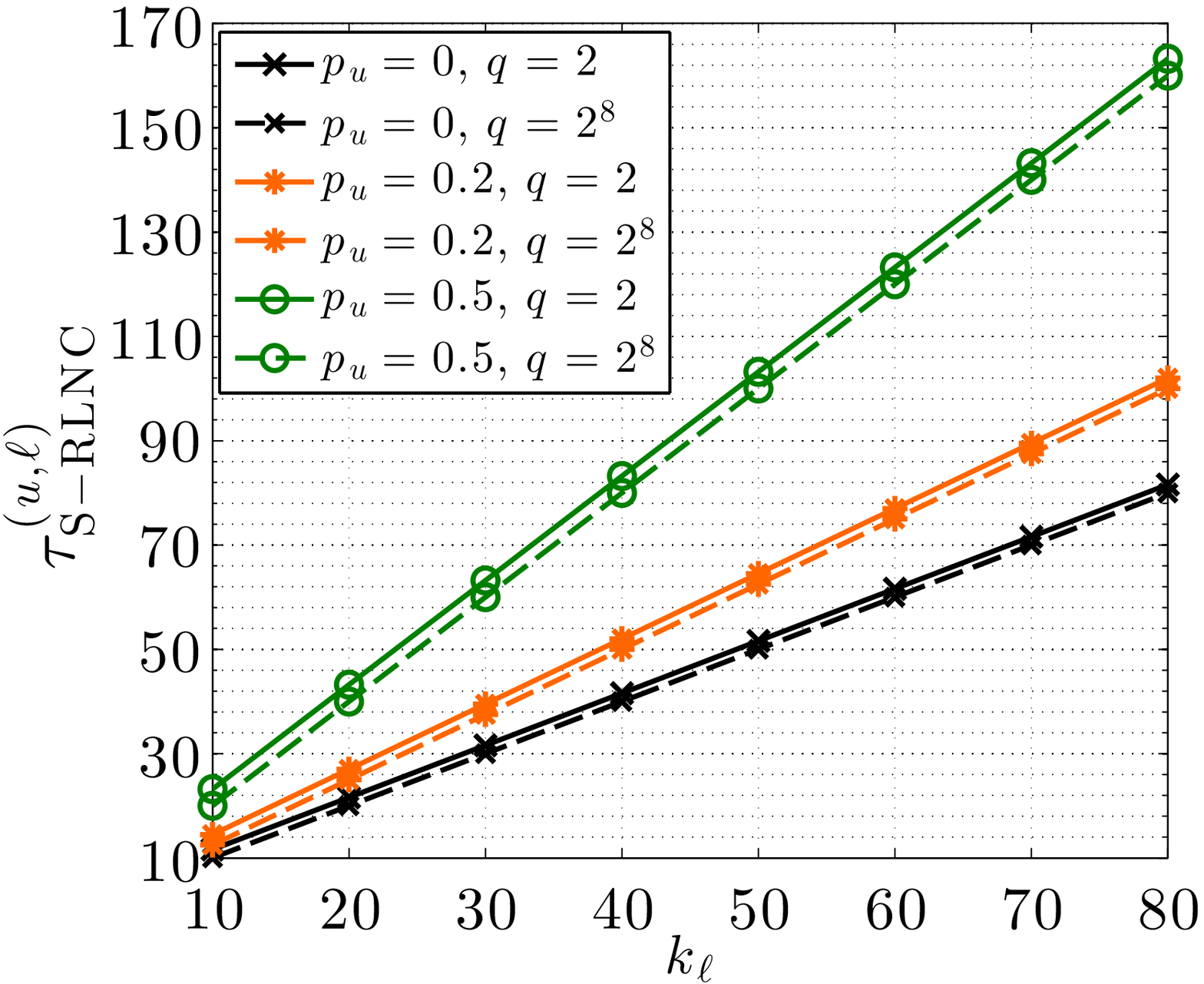}
}
\subfloat[Terms $r \cdot \phi_{\mathrm{X}}^{(u,\ell)}(r)$]{\label{fig.r2.expTerm}
\hspace{-2.1mm}\includegraphics[width=0.499\columnwidth]{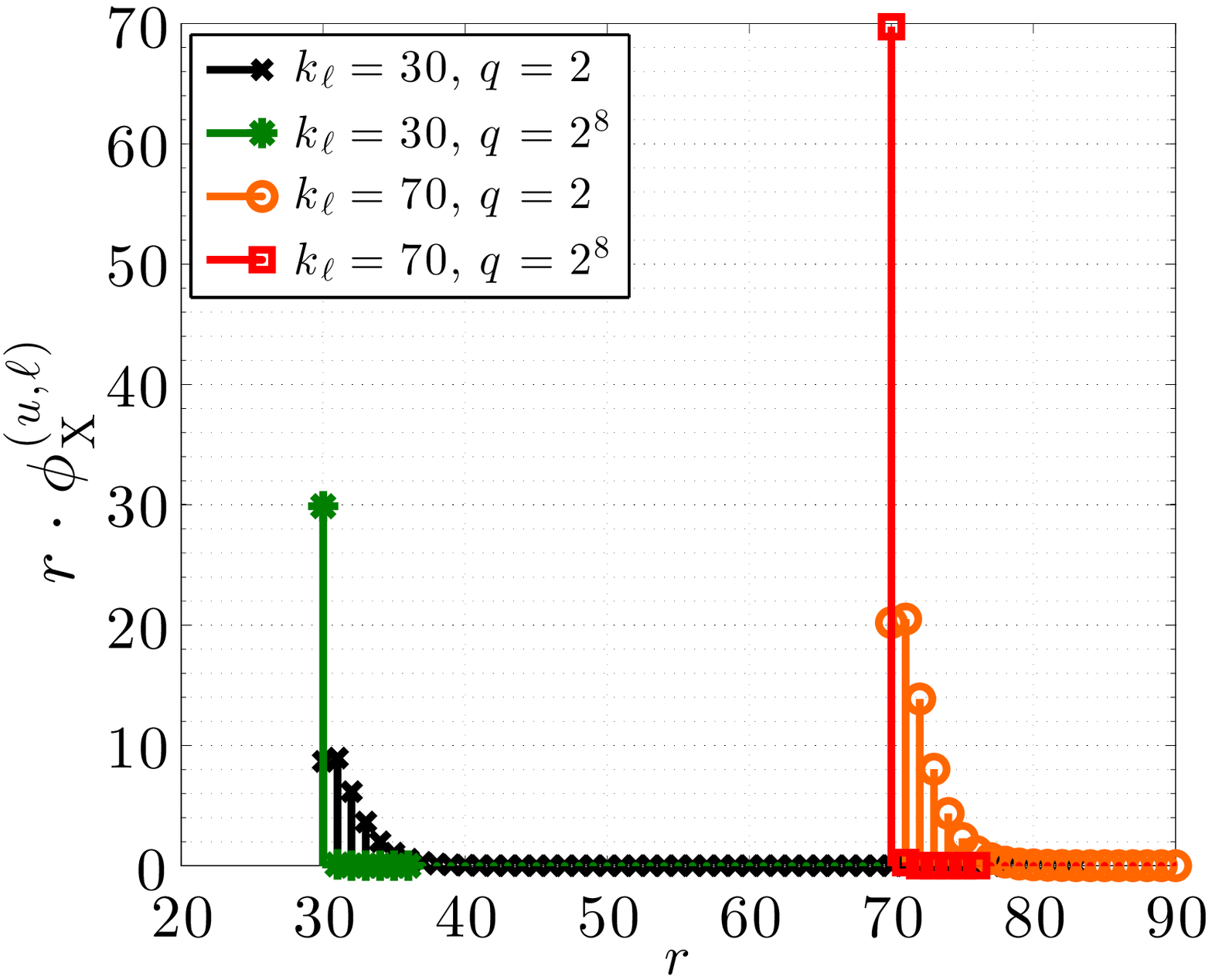}
}
\caption{Expected value and scaled PMF of $\mathrm{X}$, in the case of RLNC and for $q = 2$ and $p_\ell = 1/2$. The rightmost figure refers to case where $k_\ell = \{30, 70\}$ and $p_u = 0$.}
\label{fig.r2}
\end{figure}

From Fig.~\ref{fig.vl3.del} and regardless of the network coding strategy in use, we observe that the values of $\tau^{(u,1:\ell)}$ when $q = 2$ are very close to those obtained when $q = 2^8$. In particular, the greatest performance gap is associated in the case of RLNC and it is smaller than $6$ coded packets. In general, the performance differences between the case where $q = 2$ and the case $q = 2^8$ tend to vanish as we refer to the optimized S-RLNC or S-SRLNC strategies. The reasoning behind the aforementioned behaviour is given in the following remark.

\begin{remark}
Let $\mathrm{X}$ be a random variable expressing the number of coded packet transmissions needed to recover a message layer composed of $k_\ell$ source packets transmitted via the RLNC principle, for $p_\ell = 1/q$. Fig.~\ref{fig.r2.exp} shows, for different values of $p_u$, the expected value of $\mathrm{X}$ that is $\mathbb{E}[\mathrm{X}] = \tau_{\mathrm{S-RLNC}}^{(u,\ell)}$. In Fig.~\ref{fig.r2.exp}, we observe that the values of $\mathbb{E}[\mathrm{X}]$ derived when $q = 2$ are close to those obtained when $q = 2^8$, regardless of the value of $p_u$. For simplicity, let us refer to the case where $p_u = 0$ and $p_\ell = 1/q$. From~\cite[Eq.~(6)]{jsacTassi}, the Probability Mass Function (PMF) $\mathrm{Pr}[\mathrm{X} = r] = \phi_{\mathrm{X}}^{(u,\ell)}(r)$ of $\mathrm{X}$ can be expressed as follows:
\begin{equation}
\phi_{\mathrm{X}}^{(u,\ell)}(r) \!=\!\! \begin{cases}
			\!\displaystyle\prod_{i=0}^{k_\ell - 1} \Big[1 - q^{i-r}\Big]\text{, if $r = k_\ell$}\\
			\!\displaystyle\prod_{i=0}^{k_\ell - 1} \Big[1 - q^{i-r}\Big]\!\! -\!\! \prod_{i=0}^{k_\ell - 1} \Big[1 - q^{i-r+1}\Big]\text{, if $r > k_\ell$.}
		 \end{cases}\hspace{-10mm}\label{eq.pmf}
\end{equation}
Hence, in this case, the expected value of $\mathrm{X}$ can be alternatively expressed as $\mathbb{E}[\mathrm{X}] = \sum_{r = k_\ell}^\infty r \cdot \phi_{\mathrm{X}}^{(u,\ell)}(r)$. Fig.~\ref{fig.r2.expTerm} shows the product of terms $r \cdot \phi_{\mathrm{X}}^{(u,\ell)}(r)$ as a function of $r$, for $k_\ell = \{30, 70\}$ and $q = \{2, 2^8\}$. In the case of $q = 2$, we observe that the product $r \cdot \phi_{\mathrm{X}}^{(u,\ell)}(r)$ and hence, the PMF of $\mathrm{X}$ is non-zero across several values of $r \geq k_\ell$, for both of the considered values of $k_\ell$. On the other hand, the PMF of $\mathrm{X}$ is non-zero almost for $r = k_\ell$, when $q = 2^8$. Considering a target value of $k_\ell$ and $q = 2$, from Fig.~\ref{fig.r2.expTerm}, we can infer that the sum of non-zero terms $r \cdot \phi_{\mathrm{X}}^{(u,\ell)}(r)$, i.e., $\mathbb{E}[\mathrm{X}]$ is barely greater than or equal to the correspondent sum of terms obtained by considering $q = 2^8$. In fact, in the case of $k_\ell = 30$ ($k_\ell = 70$), $\mathbb{E}[\mathrm{X}]$ is equal to $31.6$ and $30$ ($71.6$ and $70$), for $q = 2$ and $2^8$, respectively. We thus observe that, even though the PMFs of $\mathrm{X}$ for $q=2$ and $q=2^8$ are significantly different, the corresponding average values of $\mathrm{X}$ are comparable. On the other hand, for a given $r \geq 0$, we observe that the probability value $\mathrm{Pr}[\mathrm{X} \leq r]$ may vary significantly as the value of $q$ changes. The same reasoning can be easily extended for different values of PER and applies to all the RLNC strategies discussed in this paper. In addition, that explains the reason way scenarios where $q = 2$ and $q = 2^8$ perform similarly from the point of view of the average transmission footprint.
\end{remark}

Consider Fig.~\ref{fig.vl3.del}, we observe that all the considered network coding strategies can meet the services coverage constraints, for the considered values of $\Hat{\tau}$ and $\Hat{U}_\ell$. However, since the optimized S-RLNC and S-SRLNC strategies are characterized by values of $p_\ell$ (for any $\ell = 1, \ldots, L$) that are greater than $q^{-1}$, the probability of transmitting non-degenerate coded packets associated with all-zero coding vectors is likely to increase. Hence, the average transmission footprints provided by S-RLNC and S-SRLNC are greater than those associated with the optimized RLNC and SRLNC strategies.

\begin{figure}[tb]
\centering
\subfloat[Stream A]{\label{fig.vl3.fpGain}
\includegraphics[width=0.95\columnwidth]{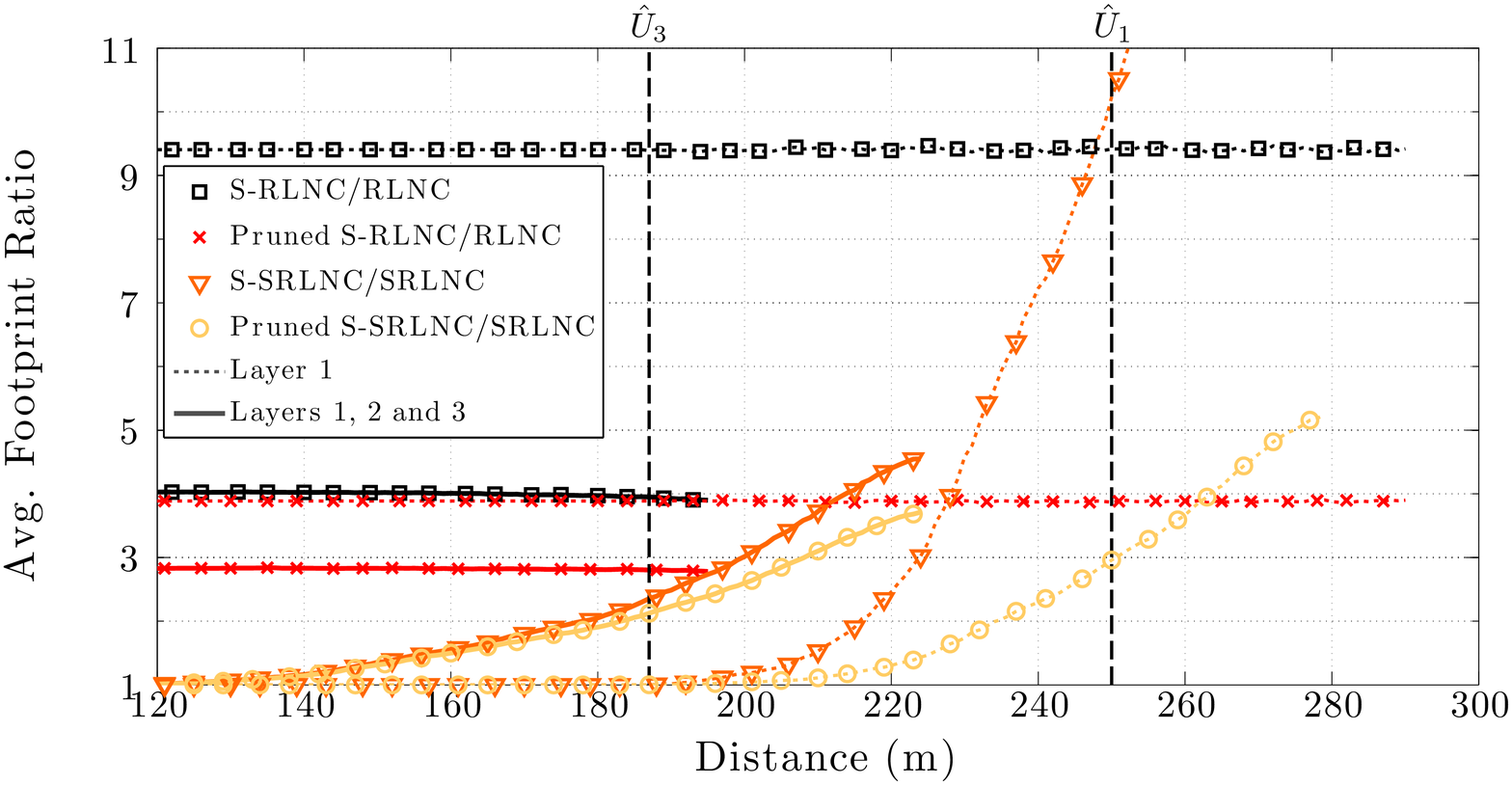}
}\\
\vspace{-3mm}\subfloat[Stream B]{\label{fig.vl4.fpGain}
\includegraphics[width=0.95\columnwidth]{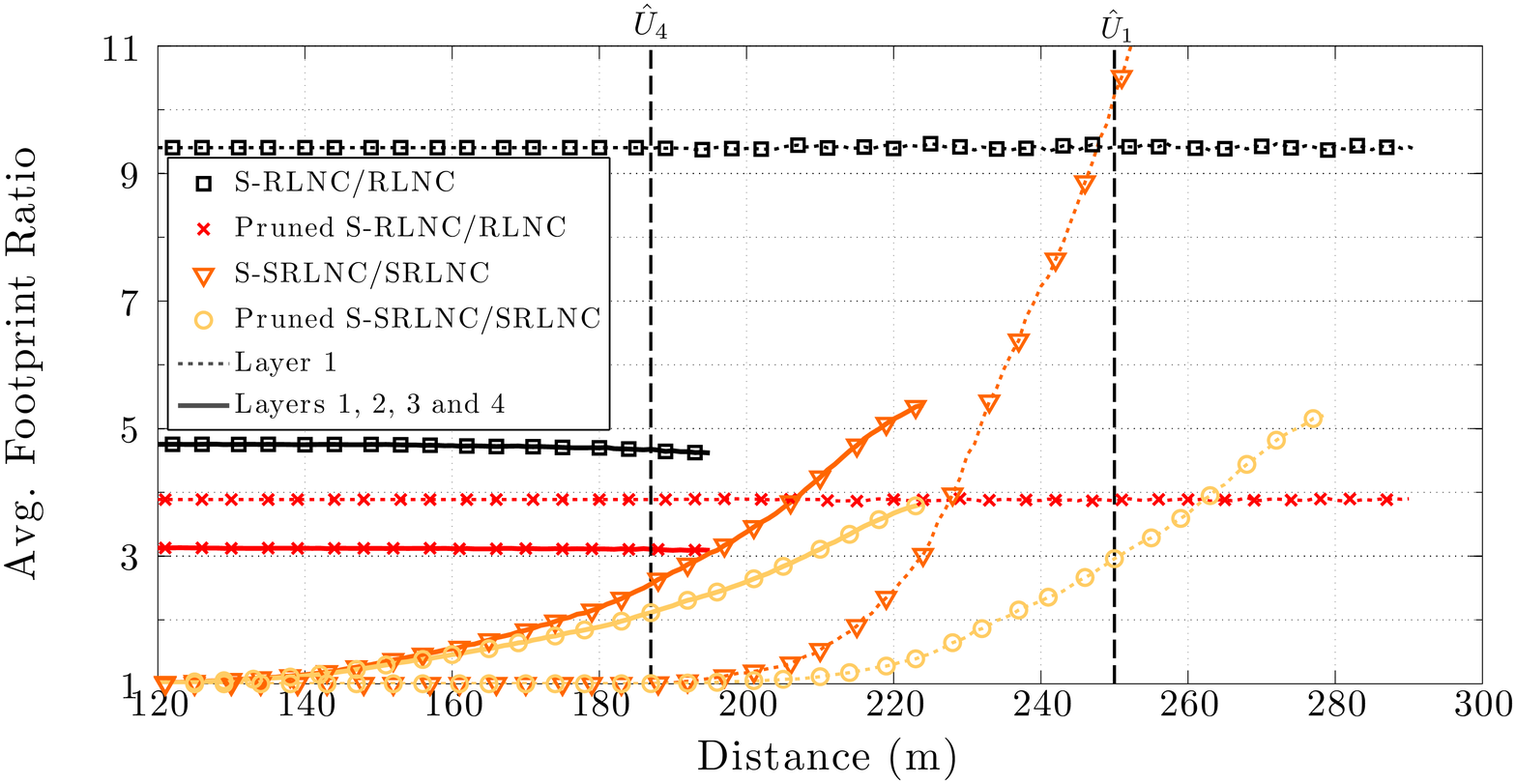}
}
\caption{Average footprint ratio $\omega^{(u,1:\ell)}$ in the case of Stream A (Stream B), for $\ell = 1$ and $3$ ($\ell = 1$ and $4$) and $q = 2$.}
\label{fig.fpGain}
\end{figure}

The aforementioned increment in the average transmission footprint has been investigated in Fig.~\ref{fig.vl3.fpGain}, where we reported the ratio $\omega^{(u,1:\ell)}$ (called ``average footprint ratio'') between the values of $\tau^{(u,1:\ell)}$ provided by the S-RLNC (\mbox{S-SRLNC}) and RLNC (SRLNC) strategies, for the QoS levels $1$ and $3$, and $q = 2$. 
We note that if $\omega^{(u,1:\ell)}$ is equal to $1$, the considered sparse network coding strategy provides the same average transmission footprint of the correspondent non-sparse technique. Fig.~\ref{fig.vl3.fpGain} also shows the same performance metrics for two modified versions of S-RLNC and S-SRLNC, hereafter referred to as ``Pruned S-RLNC'' and ``Pruned S-SRLNC''. Those two strategies behave as the proposed optimized S-RLNC and S-SRLNC but in the pruned versions, the target base station does not transmit non-degenerate coded packets associated with all-zero coding vectors. It is straightforward to prove that if a non-pruned sparse RLNC strategy meets the optimization constraints, the correspondent pruned strategy will do the same. From Fig.~\ref{fig.vl3.fpGain} we observe that the $\omega^{(u,1:\ell)}$ values provided by the S-RLNC at the target distances associated with $\Hat{U}_1$ and $\Hat{U}_3$ are equal to $9.4$ and $3.9$, respectively. However, in the case of the Pruned S-RLNC, the average footprint ratios drop to $3.9$ and $2.8$, for the QoS levels $1$ and $3$, respectively. With regards to the optimized S-SRLNC strategy, the $\omega^{(u,1:\ell)}$ values associated with the QoS level $1$ and $3$ are equal to $10.2$ and $2.4$, respectively. However, also in this case, the Pruned S-SRLNC provides smaller average footprint ratios: $2.9$ and $2.1$, for the first and the third QoS levels, respectively. 

We also observe from Fig.~\ref{fig.vl3.fpGain} that the $\omega^{(u,1:\ell)}$ values provided by (non-Pruned and Pruned) S-RLNC strategies tend to be constant, while the $\omega^{(u,1:\ell)}$ values related to the \mbox{S-SRLNC} strategies increase as the distance from the target base station grows. That behavior can be explained by the fact that non-systematic network coding strategies require to multicast coded packets from the beginning, while systematic techniques multicast coded packets only after the systematic packets have been transmitted. Hence, as the distance from the center of the cell increases, i.e., as the user propagation conditions get worse, the number of systematic packets successfully received decreases. In those cases, a user needs more coded packets to recover a video layer. However, it is worth noting that the optimized non-Pruned S-SRLNC and, specifically, the Pruned S-SRLNC strategies provide values of $\omega^{(u,1:\ell)}$ that drop below $1.6$ for distances that are $22$ m and $20$ m smaller than the desired coverage, for the QoS level $1$ and $3$, respectively. The aforementioned analysis applies also in the case of Stream B, Fig.~\ref{fig.vl4.fpGain}.

The performance of the considered network coded strategies has been also compared in terms of the complexity of the decoding operations. Likewise to the definition of $\tau^{(u,1:\ell)}$, we define the average number $\epsilon^{(1:\ell)}$ of decoding operations needed to recover the first $\ell$ video layers as follows:
\begin{equation}
\epsilon^{(1:\ell)} = \begin{cases}
			\displaystyle\sum_{t=1}^{\ell} \epsilon_{\mathrm{S-RLNC}}^{(t)}\text{,} \quad &\text{for S-RLNC}\\
			\displaystyle\sum_{t=1}^{\ell} \epsilon_{\mathrm{S-SRLNC}}^{(t)}\text{,} \quad &\text{for S-SRLNC.}
		 \end{cases}\label{eq.epsilon}
\end{equation}
Also in this case, the definition of $\epsilon^{(1:\ell)}$ can be extended to the non-sparse version of the RLNC and SRLNC, by referring to a value of $p_\ell = q^{-1}$, for $\ell = 1, \dots, L$. In the case of non-sparse RLNC and S-RLNC, we remark that the value of $\epsilon^{(1:\ell)}$ is independent from the value of the user PER. On the other hand, in the non-sparse SRLNC and S-SRLNC, the number of decoding operations grows as the number of successfully received systematic packets decreases; that happens when the user PER increases. To this end, for all the systematic strategies, we evaluated $\epsilon^{(1:\ell)}$ by referring to a user PER equal to $\Hat{p}$.

\begin{figure}[tb]
\centering
\subfloat[Stream A]{\label{fig.vl3.ops}
	\includegraphics[width=0.95\columnwidth]{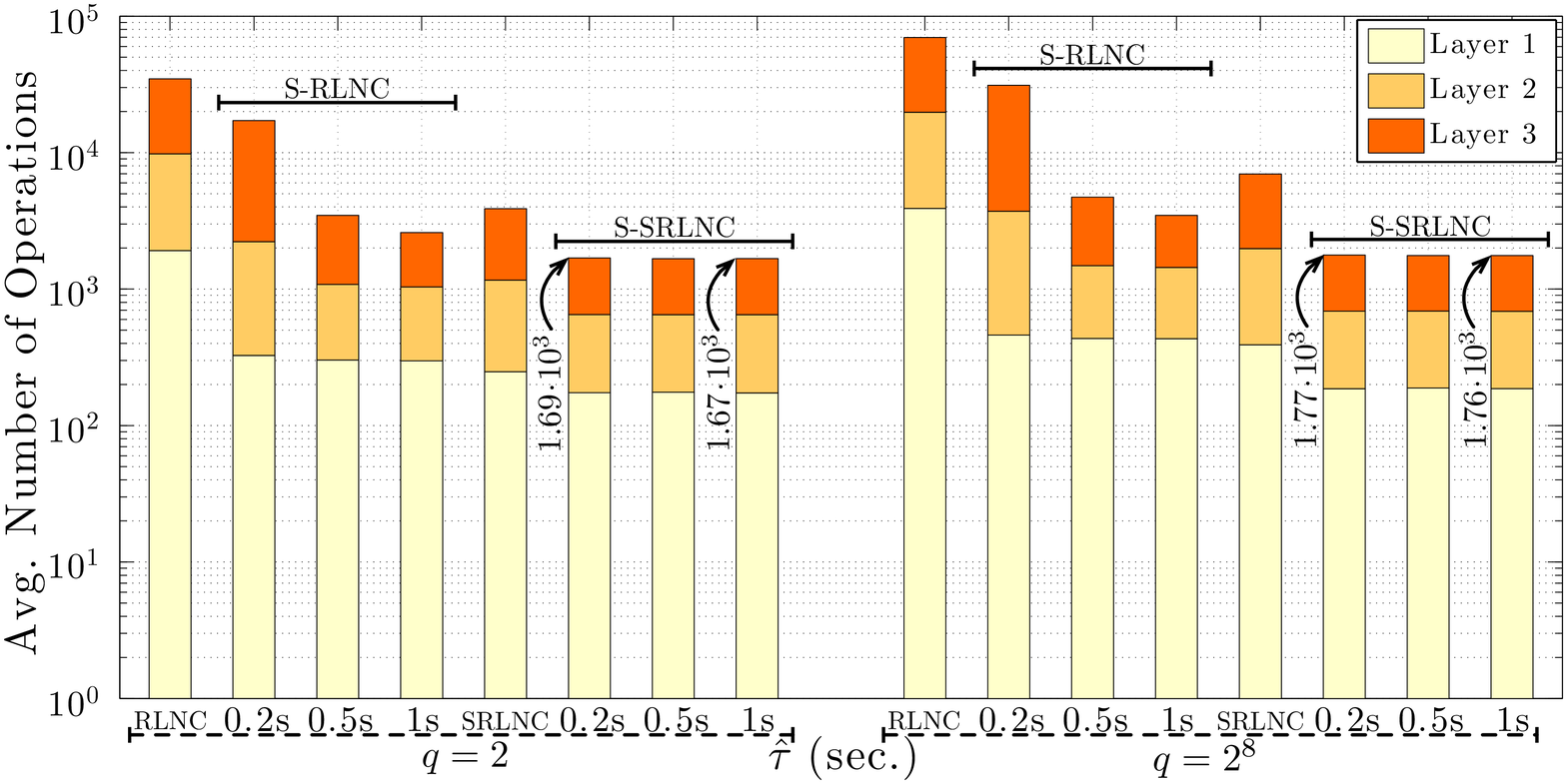}
}\\
\subfloat[Stream B]{\label{fig.vl4.ops}
	\includegraphics[width=0.95\columnwidth]{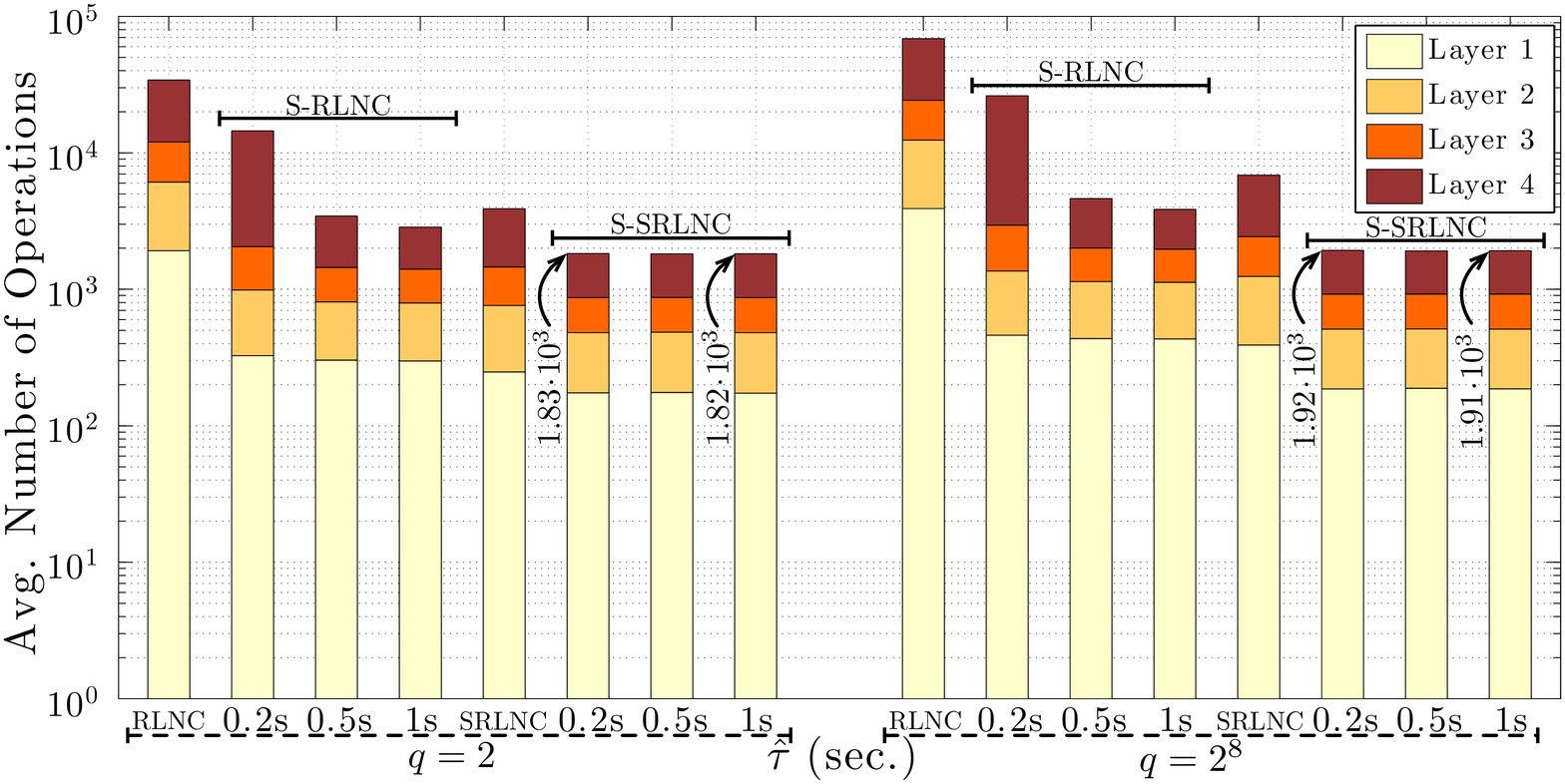}
}
\vspace{-2mm}\caption{Average number of decoding operations $\epsilon^{(1:3)}$ and $\epsilon^{(1:4)}$ in the case of Stream A and B, for $q = 2$ and $q = 2^8$. For RLNC and SRLNC, $\Hat{\tau}$ is set equal to $0.2$. Different colours represents the contribution of each layer to the value of $\epsilon^{(1:3)}$ and $\epsilon^{(1:4)}$.}
\end{figure}

Fig.~\ref{fig.vl3.ops} shows the value of $\epsilon^{(1:L)}$ provided by all the considered strategies, in the case of Stream A, for $q = \{2, 2^8\}$. We recall from Section~\ref{sec.Opt} that we refer to just the fundamental finite field operations performed by a Gaussian Elimination-based decoder. Hence, the reception of coded packets associated with all-zero coding vectors has no impact on the number of the considered operations. As a consequence, the Pruned and non-Pruned versions of S-RLNC and S-SRLNC are characterized by the same values of $\epsilon^{(1:\ell)}$.

Let us consider the S-RLNC strategy in Fig.~\ref{fig.vl3.ops}, it provides values of $\epsilon^{(1:L)}$ that are up to $92.5$\% and $97.08$\% smaller than those provided by the non-sparse RLNC, for $q = 2$ and $q = 2^8$, respectively. In particular, as expected, the value of $\epsilon^{(1:L)}$ reduces as the target service transmission time $\Hat{\tau}$ (sec.) grows. On the other hand, the S-SRLNC strategy ensures values of $\epsilon^{(1:L)}$ that are up to $57$\% and $74.8$\% smaller than those associated with the non-sparse SRLNC, for $q = 2$ and $q = 2^8$, respectively. Regardless on the value of $q$, we observe that the systematic strategies provide values of $\epsilon^{(1:L)}$ that are significantly smaller than those given by non-systematic techniques. That is due to the fact that the decoder may rely on a subset of the systematic packets that have been successfully received and do not need to be decoded. We also observe that in the case of $q = 2^8$, both the non-sparse and sparse strategies are characterized by values of $\epsilon^{(1:L)}$ that can be up to $50.2$\% greater than in the case where $q = 2$. We repeated the same performance investigation for Stream B. However, due to space limitations, we provide results only in terms of $\epsilon^{(1:L)}$, in Fig.~\ref{fig.vl4.ops}. Also in this case, the discussion and conclusions that we provided for Fig.~\ref{fig.vl3.ops} also apply.

The key points of this discussion can be summarized as follows: (i) In the considered cases, the adoption of the finite field size $q = 2^8$ provides just a minimal reduction in terms of average transmission footprint and significantly increases the complexity of the decoding operations, if compared to the $q = 2$ case, (ii) The adoption of either the Pruned \mbox{S-RLNC} or Pruned S-SRLNC ensures a significant reduction in the average number of decoding operations, (iii) The Pruned \mbox{S-SRLNC} strategy ensures the best tradeoff between transmission footprint and decoding complexity.

\section{Conclusions}\label{sec.cl}
In this paper, we addressed the issue of the complexity associated to a generic network coding decoder. In particular, we referred to a multicast network scenario where a layered service is transmitted to a set of users. 

Based on the proposed modeling, we referred to a scenario where the layered service was delivered in an ultra-reliable fashion. By referring to both the S-RLNC and S-SRLNC strategies, we proposed a constrained convex resource allocation framework suitable for jointly optimizing both the MCS indexes and the code sparsity to be used in the multicasting of each service layer. The objective of the optimization model is that of maximizing the sparsity of the code associated with each layer, and hence, minimizing the number of operations performed by a generic network coding decoder employing Gaussian Elimination. We also showed that the aforementioned computation complexity reduction can be directly mapped onto a computational processing reduction, which allows to eventually prolong the battery life of mobile devices.

As shown by the provided numerical results, the average transmission footprint is likely to increase as the sparsity of the code grows. However, the average transmission footprint can be greatly improved by simply avoiding the transmissions of coded packets associated with all-zero coding vectors, as happens with the Pruned S-RLNC and Pruned S-SRLNC strategies. We observed that the proposed optimization ensures a reduction in the average number of decoding operations of at least $92$\% and $57$\%, if compared to the classic non-sparse S-RLNC and S-SRLNC techniques, respectively. We remark that the proposed decoding complexity reduction is obtained without altering the actual implementation of the decoder.

\bibliographystyle{IEEEtran}
\bibliography{IEEEabrv,bib}

\begin{thebibliography}{10}
\providecommand{\url}[1]{#1}
\csname url@samestyle\endcsname
\providecommand{\newblock}{\relax}
\providecommand{\bibinfo}[2]{#2}
\providecommand{\BIBentrySTDinterwordspacing}{\spaceskip=0pt\relax}
\providecommand{\BIBentryALTinterwordstretchfactor}{4}
\providecommand{\BIBentryALTinterwordspacing}{\spaceskip=\fontdimen2\font plus
\BIBentryALTinterwordstretchfactor\fontdimen3\font minus
  \fontdimen4\font\relax}
\providecommand{\BIBforeignlanguage}[2]{{%
\expandafter\ifx\csname l@#1\endcsname\relax
\typeout{** WARNING: IEEEtran.bst: No hyphenation pattern has been}%
\typeout{** loaded for the language `#1'. Using the pattern for}%
\typeout{** the default language instead.}%
\else
\language=\csname l@#1\endcsname
\fi
#2}}
\providecommand{\BIBdecl}{\relax}
\BIBdecl

\bibitem{A}
P.~Popovski, ``{Ultra-Reliable Communication in 5G Wireless Systems},'' in
  \emph{Proc. of the 1st {International Conference on 5G for Ubiquitous
  Connectivity}}, Levi, Finland, FI, Nov. 2014, pp. 1--6.

\bibitem{Ai}
A.~Osseiran, F.~Boccardi, V.~Braun, K.~Kusume, P.~Marsch, M.~Maternia,
  O.~Queseth, M.~Schellmann, H.~Schotten, H.~Taoka, H.~Tullberg, M.~Uusitalo,
  B.~Timus, and M.~Fallgren, ``{Scenarios for 5G mobile and wireless
  communications: The Vision of the METIS Project},'' \emph{{IEEE} Commun.
  Mag.}, vol.~52, no.~5, pp. 26--35, May 2014.

\bibitem{6619579}
R.~Ferrus, O.~Sallent, G.~Baldini, and L.~Goratti, ``{LTE: The Technology
  Driver for Future Public Safety Communications},'' \emph{{IEEE} Commun.
  Mag.}, vol.~51, no.~10, pp. 154--161, Oct. 2013.

\bibitem{Tetra5G}
\BIBentryALTinterwordspacing
L.~Carl\'a, R.~Fantacci, F.~Gei, D.~Marabissi, and L.~Micciullo, ``{LTE}
  {enhancements for Public Safety and Security communications to support Group
  Multimedia Communications},'' \emph{{IEEE} Netw.}, to appear 2015. [Online].
  Available: \url{http://arxiv.org/abs/1501.03613}
\BIBentrySTDinterwordspacing

\bibitem{6025326}
P.~Seeling and M.~Reisslein, ``{Video Transport Evaluation With H.264 Video
  Traces},'' \emph{{IEEE} Commun. Surveys Tuts.}, vol.~14, no.~4, pp.
  1142--1165, Fourth 2012.

\bibitem{MetisAL}
J.~Monserrat, G.~Mange, V.~Braun, H.~Tullberg, G.~Zimmermann, and
  {\"O}.~Bulakci, ``{METIS Research Advances Towards the 5G Mobile and Wireless
  System Definition},'' \emph{EURASIP J. Wirel. Commun. Netw.}, vol. 2015,
  no.~1, 2015.

\bibitem{4441773}
B.-S. Kim, S.~W. Kim, and R.~Ekl, ``{OFDMA-Based Reliable Multicasting MAC
  Protocol for WLANs},'' \emph{{IEEE} Trans. Veh. Technol.}, vol.~57, no.~5,
  pp. 3136--3145, Sep. 2008.

\bibitem{KiJiKSSc10}
J.~Kim, H.~Jin, D.~K. Sung, and R.~Schober, ``{Optimization of Wireless
  Multicast Systems Employing Hybrid-ARQ with Chase Combining},'' \emph{{IEEE}
  Trans. Veh. Technol.}, vol.~59, no.~7, pp. 3342--3355, Sep. 2010.

\bibitem{6416071}
E.~Magli, M.~Wang, P.~Frossard, and A.~Markopoulou, ``{Network Coding Meets
  Multimedia: A Review},'' \emph{{IEEE} Trans. Multimedia}, vol.~15, no.~5, pp.
  1195--1212, 2013.

\bibitem{RaptorQ}
T.~Mladenov, S.~Nooshabadi, and K.~Kim, ``{Efficient GF(256) Raptor Code
  Decoding for Multimedia Broadcast/Multicast Services and Consumer
  Terminals},'' \emph{IEEE Trans. Consum. Electron.}, vol.~58, no.~2, pp.
  356--363, May 2012.

\bibitem{6353397}
C.~Khirallah, D.~Vukobratovi\'c, and J.~Thompson, ``{Performance Analysis and
  Energy Efficiency of Random Network Coding in LTE-Advanced},'' \emph{{IEEE}
  Trans. Wireless Commun.}, vol.~11, no.~12, pp. 4275--4285, Dec. 2012.

\bibitem{jsacTassi}
A.~Tassi, I.~Chatzigeorgiou, and D.~Vukobratovi\'c, ``{Resource Allocation
  Frameworks for Network-coded Layered Multimedia Multicast Services},''
  \emph{IEEE J. Sel. Areas Commun.}, vol.~33, no.~2, pp. 141--155, Feb. 2015.

\bibitem{6774596}
D.~Ferreira, R.~Costa, and J.~Barros, ``{Real-Time Network Coding for Live
  Streaming in Hyper-Dense WiFi Spaces},'' \emph{{IEEE} J. Sel. Areas Commun.},
  vol.~32, no.~4, pp. 773--781, Apr. 2014.

\bibitem{KODO}
M.~V. Pedersen, J.~Heide, and F.~H. Fitzek, ``{Kodo: An Open and Research
  Oriented Network Coding Library},'' in \emph{NETWORKING 2011 Workshops}, ser.
  Lecture Notes in Computer Science, V.~Casares-Giner, P.~Manzoni, and A.~Pont,
  Eds.\hskip 1em plus 0.5em minus 0.4em\relax Springer Berlin Heidelberg, 2011,
  vol. 6827, pp. 145--152.

\bibitem{RR1}
M.~Xiao, M.~Medard, and T.~Aulin, ``{Cross-Layer Design of Rateless Random
  Network Codes for Delay Optimization},'' \emph{{IEEE} Trans. Commun.},
  vol.~59, no.~12, pp. 3311--3322, Dec. 2011.

\bibitem{R1}
A.~Ramasubramonian and J.~Woods, ``{Video Multicast Using Network Coding},'' in
  \emph{Proc. of SPIE Conference on Visual Communications and Image Processing
  (VCIP)}, Jan. 2009, pp. 1--11.

\bibitem{R2}
H.~Wang, S.~Xiao, and C.-C.~J. Kuo, ``{Random Linear Network Coding with
  Ladder-shaped Global Coding Matrix for Robust Video Transmission},''
  \emph{Elsevier J. Vis. Comun. Image Represent.}, vol.~22, no.~3, pp.
  203--212, Apr. 2011.

\bibitem{R3}
N.~Thomos, J.~Chakareski, and P.~Frossard, ``{Prioritized Distributed Video
  Delivery With Randomized Network Coding},'' \emph{{IEEE} Trans. Multimedia},
  vol.~13, no.~4, pp. 776--787, Aug. 2011.

\bibitem{6849990}
A.~Tassi, C.~Khirallah, D.~Vukobratovi\'c, F.~Chiti, J.~Thompson, and
  R.~Fantacci, ``{Resource Allocation Strategies for Network-Coded Video
  Broadcasting Services over LTE-Advanced},'' \emph{{IEEE} Trans. Veh.
  Technol.}, vol.~64, no.~5, pp. 2186--2192, May 2015.

\bibitem{RR3}
L.~Lu, M.~Xiao, M.~Skoglund, L.~Rasmussen, G.~Wu, and S.~Li, ``{Efficient
  Network Coding for Wireless Broadcasting},'' in \emph{Proc. of IEEE WCNC
  2010}, Sydney, Australia, AUS, Apr. 2010, pp. 1--6.

\bibitem{RR2}
L.~Lu, M.~Xiao, and L.~Rasmussen, ``{Relay-Aided Broadcasting with
  Instantaneously Decodable Binary Network Codes},'' in \emph{Proc. of ICCCN
  2011}, Maui, Hawaii, HI, Jul. 2011, pp. 1--5.

\bibitem{LucaniNetCod}
S.~Feizi, D.~Lucani, C.~S{\o}rensen, A.~Makhdoumi, and M.~M\'edard, ``{Tunable
  Sparse Network Coding for Multicast Networks},'' in \emph{Proc. of NetCod
  2014}, Aalborg, Denmark, DK, Jun. 2014.

\bibitem{ComplexityGE}
D.~Andr\'en, L.~Hellstr{\o}m, and K.~Markstr{\o}m, ``{On the Complexity of
  Matrix Reduction Over Finite Fields},'' \emph{Advances in Applied
  Mathematics}, vol.~39, no.~4, pp. 428--452, 2007.

\bibitem{5061923}
J.~Barros, R.~Costa, D.~Munaretto, and J.~Widmer, ``{Effective Delay Control in
  Online Network Coding},'' in \emph{Proc. of IEEE INFOCOM 2009}, Rio de
  Janeiro, Brazil, BR, Apr. 2009, pp. 208--216.

\bibitem{5723187}
R.~Prior and A.~Rodrigues, ``{Systematic Network Coding for Packet Loss
  Concealment in Broadcast Distribution},'' in \emph{Proc. of ICOIN 2011}, Jan.
  2011, pp. 245--250.

\bibitem{ICicc2015}
A.~Jones, I.~Chatzigeorgiou, and A.~Tassi, ``{Binary Systematic Network Coding
  for Progressive Packet Decoding},'' in \emph{Proc. of IEEE ICC 2015}, London,
  United Kingdom, UK, Jun. 2015, pp. 1--5.

\bibitem{5963470}
X.~Li, W.~H. Mow, and F.-L. Tsang, ``{Singularity Probability Analysis for
  Sparse Random Linear Network Coding},'' in \emph{Proc. of IEEE ICC 2011},
  Kyoto, Japan, JP, Jun. 2011, pp. 1--5.

\bibitem{5978939}
------, ``{Rank Distribution Analysis for Sparse Random Linear Network
  Coding},'' in \emph{Proc. of NetCod 2011}, Beijing, China, CN, Jul. 2011, pp.
  1--6.

\bibitem{RSA:RSA1}
J.~Bl\"{o}mer, R.~Karp, and E.~Welzl, ``{The Rank of Sparse Random Matrices
  Over Finite Fields},'' \emph{Random Structures \& Algorithms}, vol.~10,
  no.~4, pp. 407--419, 1997.

\bibitem{Cooper2000}
C.~Cooper, ``{On the Asymptotic Distribution of Rank of Random Matrices Over a
  Finite Field},'' \emph{Random Structures \& Algorithms}, vol.~17, no. 3-4,
  pp. 197--212, 2000.

\bibitem{ericsson}
\BIBentryALTinterwordspacing
{ERICSSON AB}, ``{5G Radio Access},'' Tech. Rep., Feb. 2015. [Online].
  Available: \url{http://www.ericsson.com/res/docs/whitepapers/wp-5g.pdf}
\BIBentrySTDinterwordspacing

\bibitem{sesia2011lte}
S.~Sesia, I.~Toufik, and M.~Baker, \emph{{LTE - The UMTS Long Term
  Evolution}}.\hskip 1em plus 0.5em minus 0.4em\relax John Wiley \& Sons, 2011.

\bibitem{h264}
{ITU-T H.264}, ``{Advanced Video Coding for Generic Audiovisual Services},''
  Tech. Rep., Nov. 2007.

\bibitem{6148193}
R.~Afolabi, A.~Dadlani, and K.~Kim, ``{Multicast Scheduling and Resource
  Allocation Algorithms for OFDMA-Based Systems: A Survey},'' \emph{{IEEE}
  Commun. Surveys Tuts.}, vol.~15, no.~1, pp. 240--254, First 2013.

\bibitem{4459059}
C.~Fragouli, J.~Widmer, and J.-Y. Le~Boudec, ``{Efficient Broadcasting Using
  Network Coding},'' \emph{IEEE/ACM Trans. Netw.}, vol.~16, no.~2, pp.
  450--463, Apr. 2008.

\bibitem{6991518}
A.~Cohen, B.~Haeupler, C.~Avin, and M.~M\'edard, ``{Network Coding Based
  Information Spreading in Dynamic Networks With Correlated Data},'' \emph{IEEE
  J. Sel. Areas Commun.}, vol.~33, no.~2, pp. 213--224, Feb. 2015.

\bibitem{5634159}
O.~Trullols-Cruces, J.~Barcelo-Ordinas, and M.~Fiore, ``{Exact Decoding
  Probability Under Random Linear Network Coding},'' \emph{{IEEE} Commun.
  Lett.}, vol.~15, no.~1, pp. 67--69, Jan. 2011.

\bibitem{FMC60}
J.~Kem{\'e}ny and J.~Snell, \emph{Finite Markov Chains}.\hskip 1em plus 0.5em
  minus 0.4em\relax Van Nostrand, 1960.

\bibitem{upton2014raspberry}
E.~Upton and G.~Halfacree, \emph{Raspberry Pi User Guide}.\hskip 1em plus 0.5em
  minus 0.4em\relax John Wiley \& Sons, 2014.

\bibitem{Boyd:2004:CO:993483}
S.~Boyd and L.~Vandenberghe, \emph{Convex Optimization}.\hskip 1em plus 0.5em
  minus 0.4em\relax New York, NY, USA: Cambridge University Press, 2004.

\bibitem{TR_36_814}
{3GPP TR 36.814 v9.0.0}, ``{Evolved Universal Terrestrial Radio Access
  (E-UTRA); Further advancements for E-UTRA physical layer aspects},'' 2010.

\bibitem{StreamA}
\BIBentryALTinterwordspacing
``{Video Traces 2 for Terminator (DQP 15, 2 EL) and (DQP 10, 3 EL)}.''
  [Online]. Available: \url{http://trace.eas.asu.edu/videotraces2/cgs/}
\BIBentrySTDinterwordspacing

\end{thebibliography}

\end{document}